**InAs on Insulator: A New Platform for Cryogenic Hybrid Superconducting Electronics**

*Alessandro Paghi*[*], *Giacomo Trupiano, Giorgio De Simoni, Omer Arif, Lucia Sorba, and Francesco Giazotto*

Istituto Nanoscienze-CNR and Scuola Normale Superiore, Piazza San Silvestro 12, 56127 Pisa, Italy.

[*]Corresponding authors: alessandro.paghi@nano.cnr.it

## Abstract

Superconducting circuits based on hybrid InAs Josephson Junctions (JJs) play a starring role in the design of fast and ultra-low power consumption solid-state quantum electronics and exploring novel physical phenomena. Conventionally, 3D substrates, 2D quantum wells (QWs), and 1D nanowires (NWs) made of InAs are employed to create superconducting circuits with hybrid JJs. Each platform has its advantages and disadvantages. Here, we proposed the InAs-on-insulator (InAsOI) as a groundbreaking platform for developing superconducting electronics. An epilayer of semiconducting InAs with different electron densities was grown onto an InAlAs metamorphic buffer layer, efficiently used as a cryogenic insulator to decouple adjacent devices electrically. JJs with various lengths and widths were fabricated employing Al as a superconductor and InAs with different electron densities. We achieved a switching current density of 7.3 μA/μm, a critical voltage of 50-to-80 μV, and a critical temperature equal to that of the superconductor used. For all the JJs, the switching current follows a characteristic Fraunhofer pattern with an out-of-plane magnetic field. These achievements enable the use of InAsOI to design and fabricate surface-exposed Josephson Field Effect Transistors with high critical current densities and superior gating properties.

## Introduction

Superconducting circuits based on Josephson Junctions (JJs) play a starring role in the design of fast and ultra-low power consumption solid-state quantum electronics [1][2]. Over the last decades, hybrid superconductor-semiconductor JJs have attracted growing interest as basic blocks to explore novel physical phenomena [3][4][5][6] and to build quantum electronic architectures [7][8]. Various applications have been documented, encompassing gate-tunable superconducting [9][10][11][12] and Andreev [13][14] qubits, superconducting transistors [15][16][17], diodes [18][19][20], and interferometers [21][22][23]. These applications extend to quantum phenomena such as spin-dependent supercurrent [24], topological phase transitions [25][26][27], anomalous phase shifts in ground state [28][29], and parity-protected systems [30][31][32].
In the III-V compound group, InAs is a semiconductor with a natural surface charge accumulation that pins the Fermi-level above its conduction band edge [33][34], which allows the formation of Ohmic contacts with conventional metals. Furthermore, when superconducting metals are employed, the superconductive properties can be lent to the InAs layer via the superconducting proximity effect [35][36]. Among InAs-based platforms, 3D substrates, 2D quantum wells (QWs), and 1D nanowires (NWs) are conventionally employed to implement circuits featuring hybrid superconductor-semiconductor JJs.

JJs and gate-tunable JJs, also known as Josephson Field Effect Transistors (JoFETs), were initially reported in 3D InAs substrates featuring surface 2-dimensional electron gases (2DEGs) [37][38][39]. A notable critical current density ($I_C/W$, where W represents the junction width) of 20 µA/µm was obtained involving Nb electrodes (Figure 1a, blue dots, Table S1), although practical implementations of this platform face obstacles due to the presence of an InAs conductive path between adjacent devices [37].

Buried [15][40] and newest near-surface [16][41][42][43][44] InAs 2D QWs are today the most used platform to host high-mobility 2DEGs. Despite the material-related improvements of recent years, Al-contacted InAs QWs always feature critical current densities at least 1-to-2 orders of magnitude lower than what is achieved with InAs 3D substrates (Figure 1a green dots, Table S1).

InAs 1D NWs have also been the object of an intense research effort due to their potential applications as building blocks in nanoscale quantum devices [45][9][46]. As the InAs 2D and 3D counterparts, InAs NWs support superconductivity via the superconducting proximity effect [47][26][48]. Regardless of their 1D morphology, Al-proximitized InAs NWs exhibit critical current densities similar to those achieved in QWs, with the exception of ballistic NWs proximitized with Al [47] or Pb [49] leads, where an Ic/W of 10 μA/μm was achieved (Figure 1a black dots, Table S1). Despite electrical performance, real applications of InAs NWs-based devices are limited by the NW placing in specific locations on the surface with nanometric precision [50][51][52][53].

Here, we propose the InAs-on-insulator (InAsOI) as a new platform to develop planar superconductive electronics. A 100-nm-thick epilayer of semiconductive InAs with different doping levels was grown onto an InAlAs metamorphic buffer. In our scheme, the metamorphic buffer is efficiently used as a cryogenic insulator to decouple adjacent devices electrically. JJs with different lengths and widths were fabricated employing Al as a superconductor and InAs with different electron densities. The critical current density is as high as 7.3 μA/μm for a JJ with a length of 350 nm and InAs sheet electron density of $10^{14}$ $cm^{-2}$, comparable to InAs 3D substrates. We emphasize that InAsOI resembles the highly successful silicon-on-insulator (SOI) architecture and constitutes a promising candidate for implementing the superconducting counterpart of classic semiconductive electronics made on SOI.

## Results and Discussion

The InAsOI platform for cryogenic superconducting electronics is shown in Figure 1b.

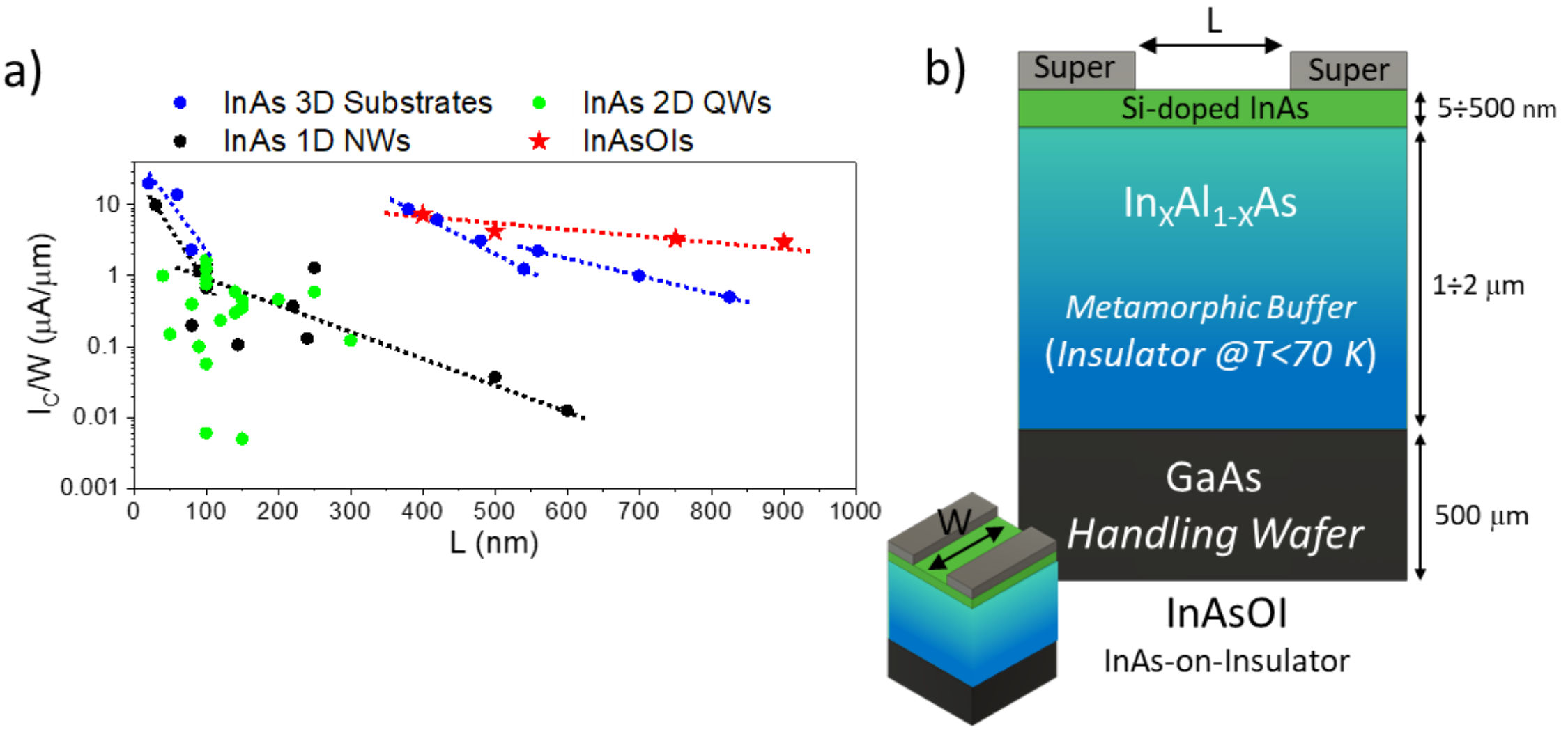

**Figure 1: InAs-on-insulator for cryogenic superconducting electronics: concept and comparison.** a) State-of-art critical current densities achieved with InAs-based superconducting platforms. The most representative devices are indicated with a dot for each platform, namely InAsOIs, InAs 3D substrates, InAs 2D QWs, and InAs 1D NWs. Where available, we compared InAs-platforms using Al as a superconductor. InAsOIs exhibit critical current densities higher than the competitors (comparable with those achieved with InAs 3D substrates + Nb leads). These are also tunable by orders of magnitude by choosing the InAs sheet electron density. b) InAsOI in-section and 3D platform structure.

The surface-exposed InAs epilayer supports a non-dissipative current via the superconducting-proximity-effect inherited by the superconducting leads, e.g., Al [17][43], Pb [49], Nb [37][15][54], NbTi [55], and others [56][57]. The presence of InAs on the top of the structure allows direct access to the final semiconductive channel, where its electronic transport properties can be tuned, changing the InAs doping during (via atom incorporation) or post (via dopant implantation or thermal diffusion) the heterostructure growth. The μm-thick InAlAs metamorphic buffer avoids lattice mismatch between the InAs and the GaAs handling substrate and can be employed as an insulator at cryogenic temperature. Regardless of the fabrication technique, which is typical of GaAs-based heterostructures, the InAsOI represents the superconducting twin of the silicon-on-insulator (SOI) architecture, where proximitized InAs and the cryogenic insulating InAlAs replace silicon and silicon oxide layers of SOI, respectively.

Figure 2a shows details of the InAsOI heterostructure grown by molecular beam epitaxy (MBE). Despite the use of MBE, we believe that InAsOI can be fabricated also using CVD-growth techniques to speed up the wafer-scale manufacturing process without a significant worsening of transport and superconducting properties.

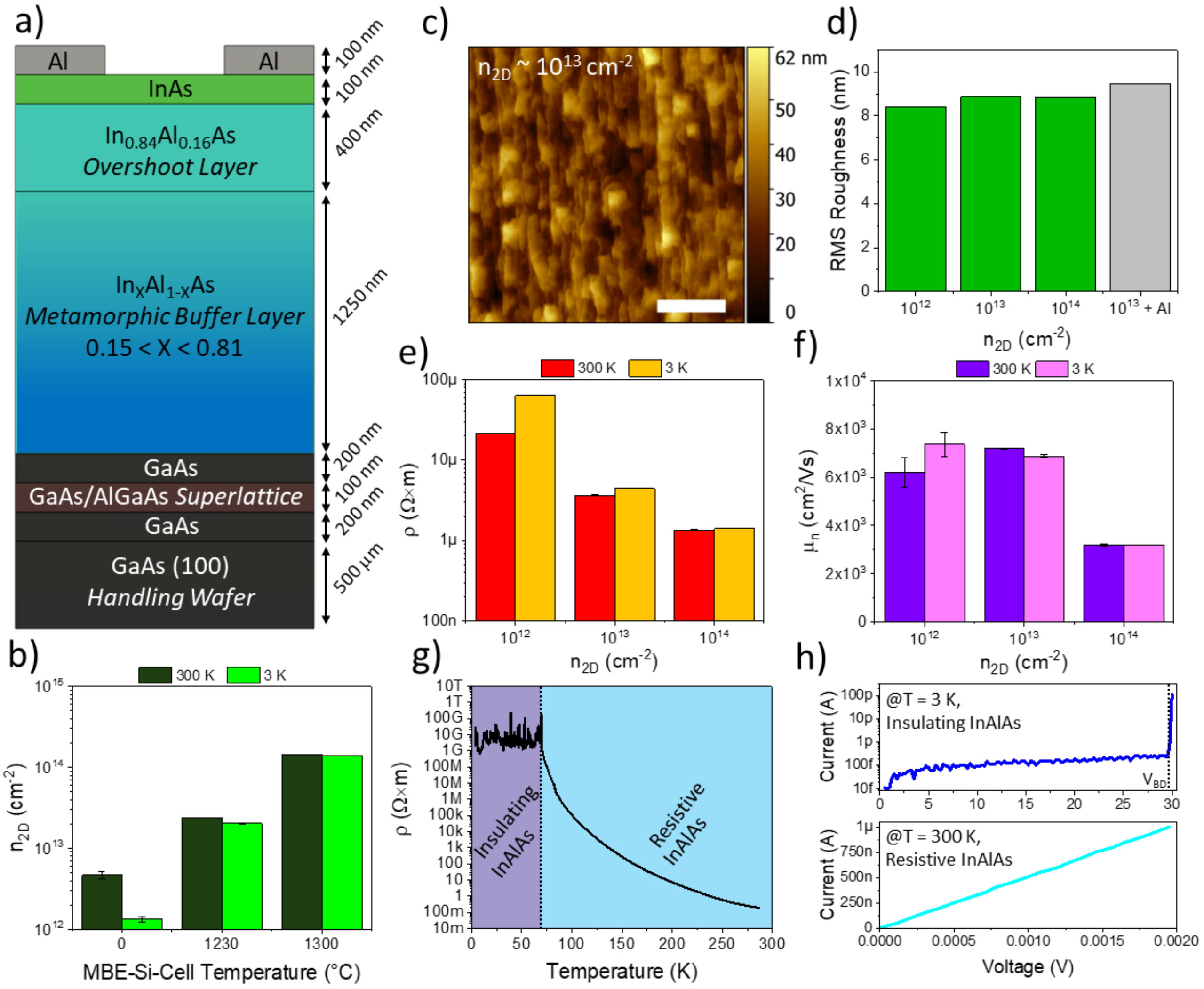

**Figure 2: InAs-on-insulator morphological and electrical characterization.** a) Cross-section layer structure of InAsOI. b) Sheet electron density of InAsOI at 3 K and 300 K vs MBE Si cell temperature (*T*=0 °C means no Si atoms). c) Atomic force microscopy photograph of an InAsOI substrate featuring $n_{2D} \sim 10^{13}$ cm$^{-2}$; the scalebar is 5 μm. d) InAs RMS roughness extrapolated from atomic force microscopy images vs. InAsOI sheet electron density. e,f) Resistivity (e) and mobility (f) of InAsOI at 3 K and 300 K vs. InAsOI sheet electron density. g) Resistivity of InAlAs metamorphic buffer vs. temperature, highlighting the insulating behavior for temperatures lower than 70 K and the resistive behavior for temperatures higher than 70 K. h) Current vs voltage curve of InAlAs metamorphic buffer at 3 K (up) and 300 K (bottom) for an InAlAs strip, highlighting the insulating behavior at 3 K and resistive behavior at 300 K. Data in b,e,f) are reported as the average value measured over n=3 samples for each InAsOI sheet electron density, with error bars representing the standard deviation.

From bottom to top, the stack consists of a 500 μm-thick semi-insulating GaAs (100) substrate, a 200 nm-thick GaAs layer, a 200 nm-thick GaAs/AlGaAs superlattice, a 200 nm-thick GaAs layer, a 1.250 μm-thick step-graded $In_XAl_{1-X}As$ metamorphic buffer with X increasing from 0.15 to 0.81, a 400 nm-thick $In_{0.84}Al_{0.16}As$ overshoot layer, and finally, a 100 nm-thick InAs epilayer. The GaAs layers and the GaAs/AlGaAs superlattice below the $In_XAl_{1-X}As$ buffer are grown to planarize the starting GaAs surface and to reduce surface roughness caused by the oxide desorption process. The graded metamorphic buffer and overshoot layer were necessary to avoid lattice mismatch between GaAs and InAs [58]. *N*-type Si doping of the InAs layer was performed by setting different MBE Si cell temperatures, namely, 0 °C (i.e., no Si doping), 1230 °C, and 1300 °C. Figure 2b shows the relationship between the MBE Si cell temperature and the InAs sheet electron density ($n_{2D}$), evaluated both at 300 K and 3 K. At 3K, the transport properties of InAsOI are totally related to the InAs layer due to the insulating behavior of the InAlAs overshoot and metamorphic buffer. We observed a constant reduction of $n_{2D}$ of ~$3.5\times10^{12}$ $cm^{-2}$ from 300 K to 3K independent of the sheet electron density value, which is related to the charge freeze-out both in the InAs and InAlAs layers. InAs $n_{2D}$ values are $1\times10^{12}$ $cm^{-2}$ (intrinsically doped), $2\times10^{13}$ $cm^{-2}$, and $1\times10^{14}$ $cm^{-2}$, respectively for Si cell temperatures of 0 °C (no Si doping), 1230 °C, and 1330 °C. The slight relative reduction of $n_{2D}$ with the temperature for Si-doped InAs epilayers indicates that donors are practically always fully ionized, which is also consistent with what is observed for Si-doped InAsNWs [59]. On the other hand, the InAsOI sheet electron density reached without Si doping could be also related to deep donor levels in the InAlAs band gap [60]. Figure 2c shows an atomic force microscopy (AFM) image of an InAsOI substrate with $n_{2D}$ ~ $10^{13}$ $cm^{-2}$, while AFM photographs of other substrates are reported in Figure S1. A surface RMS roughness of ~ 8 nm was evaluated regardless of the sheet electron concentration, which increased to ~9 nm after deposition of the 100 nm-thick Al film used as a superconductor. Figure 2e,f show InAsOI resistivity ($\rho$) and mobility ($\mu_n$) evaluated at 300 K and 3 K by Hall measurements. As reported for $n_{2D}$, a similar temperature-dependent trend was also observed for $\rho$. The cryogenic InAs resistivity decreases from ~ 60 $\mu\Omega\times$m to 1 $\mu\Omega\times$m increasing the sheet electron density from $10^{12}$ $cm^{-2}$ to $10^{14}$ $cm^{-2}$. Similarly, the cryogenic InAs mobility decreases from ~ $7\times10^3$ $cm^2$/Vs to $3\times10^3$ $cm^2$/Vs increasing $n_{2D}$ of two orders of magnitude, which are 1.5-to-5 times lower than the peak electron mobilities usually obtained from InAs near-surface 2D QWs employed in superconducting electronics (Table S1). We estimated an electron mean free path of 112 nm, 177 nm, and 156 nm in the case of InAs featuring $n_{2D}$ ~ $10^{12}$ $cm^{-2}$, $10^{13}$ $cm^{-2}$, and $10^{14}$ $cm^{-2}$, respectively. We also evaluated the contact resistance ($R_C$) between Al and InAs at 300 K and 3 K employing the Transfer Length Method (TLM), which is a standard way used to evaluate the quality of the contact between two conductors [61][62]. The contact between InAs and Al was optimized by surface removal of the InAs native oxide ($InAsO_X$) and concomitant surface Sulfur passivation, which allowed to decrease $R_C$, leaving the InAs resistivity practically unaffected (Figure S2) [63][64]. The $InAsO_X$ etching enhances the Al superconducting proximity effect, thereby increasing the critical current of the resulting Al-InAs JJs [37]. A contact resistance of ~ 6 Ω was obtained at 3 K, independent of the InAs electrical transport properties (Figure S3).

We then evaluated the temperature-dependent electrical behavior of the InAlAs metamorphic buffer by measuring the layer resistivity. As shown in Figure 2g, ρ increases from ~ 100 mΩ×m at 300 K to 10 GΩ×m at 70 K, below which it remains unchanged. Figure 2h shows current vs. voltage (I-V) curves of an InAlAs strip (200 μm width and 5 μm length) recorded both at 300 K and 3 K. At room temperature, the InAlAs exhibits a resistive behavior (Figure S5) with a linear I-V characteristic and a resistance of ~ 2 kΩ. On the other hand, at 3 K, the metamorphic buffer shows an insulating behavior with an apparent breakdown voltage ($V_{BD}$) of 30 V and a parallel resistance before the avalanche breakdown of 140 TΩ. The latter result indicates that the metamorphic buffer can be efficiently used as an insulator for temperatures lower than 70 K, which is well above every BCS superconductor critical temperature ($T_C$).

We realized superconductor-semiconductor-superconductor Al-InAs-Al JJs with several widths (W = 5, 10, 20 μm) and inter-electrode separations (length, L = 350, 500, 700, 900 nm), using InAsOIs with different sheet electron densities. Devices were fabricated via two aligned lithographic steps: first, Al and InAs MESA were defined by UV-lithography and manufactured by successive Al and InAs wet etching, setting the JJ width and leaving the cryogenic-dielectric-InAlAs layer exposed. Then, the JJ length was defined by electron-beam-lithography and Al wet-etching, leaving the underneath InAs unaffected. Figure 3a top shows an optical microscopy image of a JJ featuring a width of 20 μm and a length of 900 nm (chosen to emphasize the interelectrode separation). In contrast, the inset shows the whole device. A clear separation between the light-grey area (Al-area) and the dark-gray area (MESA etched area) is observed in the picture, with straight edges related to fine-controlled etching during device processing.

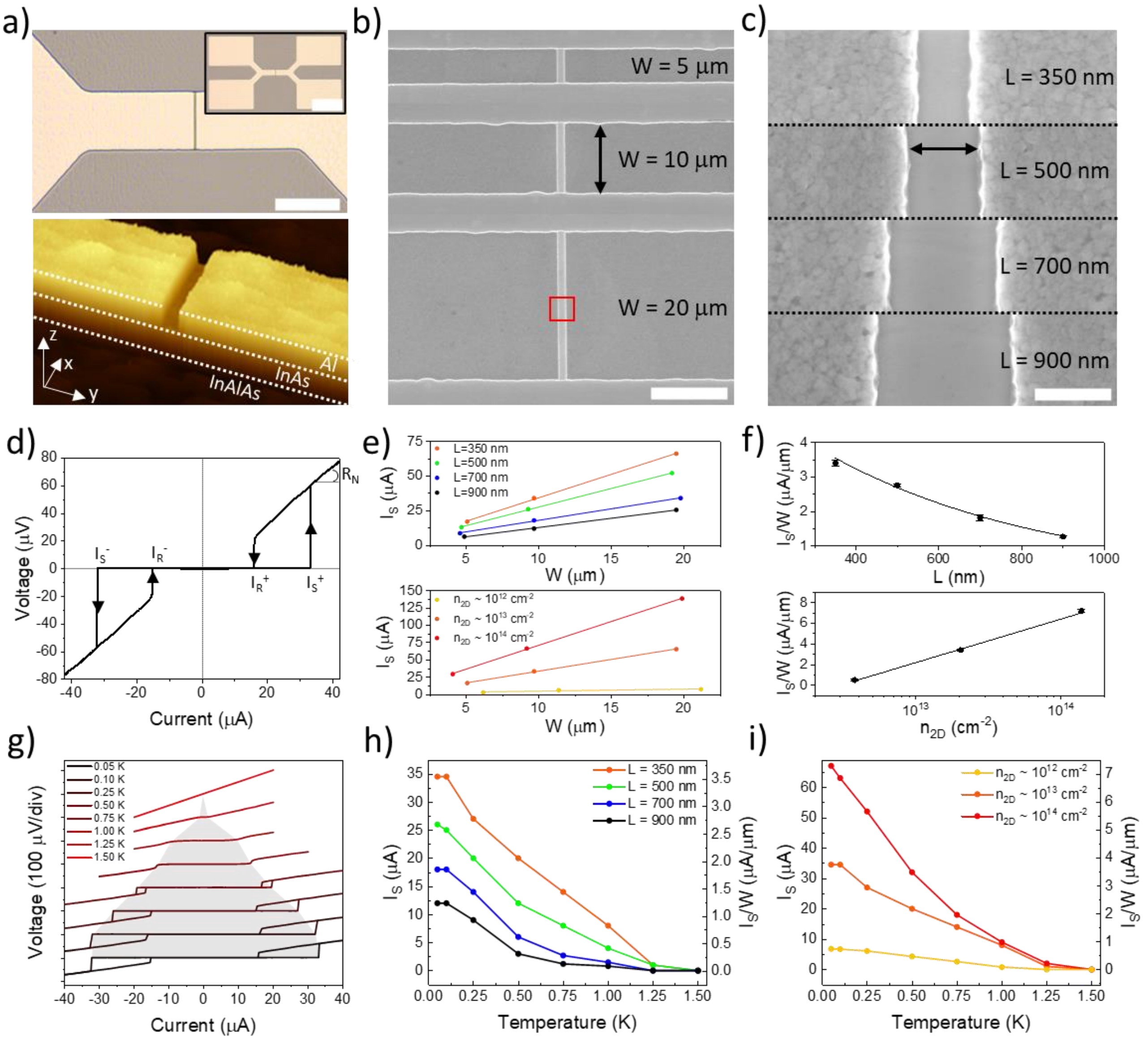

**Figure 3: InAsOI-based Josephson Junctions morphological and electrical characterization.** a) (top) Optical microscope image of a JJ featuring W = 20 μm and L = 900 nm; the scalebar is 25 μm. The inset shows the overall device; the scalebar is 100 μm. (bottom) xy-plane tilted atomic force microscopy z-profile of a JJ featuring W = 5 μm and L = 900 nm. b) Top-view scanning electron microscopy images (2.5k ×) of JJs featuring L = 900 nm and different widths; the scalebar is 10 μm. c) Top-view scanning electron microscopy images (50k ×) of JJs featuring W = 5 μm and different lengths; the scalebar is 500 nm. d) Forward and backward voltage vs- current characteristic of a JJ featuring W = 10 μm, L = 350 nm, and InAs $n_{2D}$ ~ $10^{13}$ cm$^{-2}$, measured at 50 mK. e) Switching current vs width for different lengths (top) and InAs sheet electron densities (bottom) measured at 50 mK. (top) JJs feature InAs $n_{2D}$ ~ $10^{13}$ cm$^{-2}$; (bottom) JJs feature L = 350 nm. f) Switching current density vs length (top) and InAs sheet electron density (bottom) measured at 50 mK. (top) JJs feature InAs $n_{2D}$ ~ $10^{13}$ cm$^{-2}$; (bottom) JJs feature L = 350 nm. g) Forward and backward voltage vs. current characteristics of a JJ featuring W = 10 μm, L = 350 nm, and InAs $n_{2D}$ ~ $10^{13}$ cm$^{-2}$, measured changing temperature from 0.05 to 1.5 K. h) Switching current and switching current density vs temperature for different lengths. JJs feature InAs $n_{2D}$ ~ $10^{13}$ cm$^{-2}$ and W = 10 μm. i) Switching current and switching current density vs temperature for different InAs sheet electron densities. JJs feature L = 350 nm and and W = 10 μm. Data in f) are reported as the average value measured over n=3 samples for each JJ length and InAs sheet electron density, with error bars representing the standard deviation.

The JJ shape can be seen in Figure 3a bottom, which shows the z-axis tilted profile of the junction surface (xy-plane) obtained through an AFM scan of the device (see also Figure S5a). The JJ outlines in the xz and yz planes are shown in Figure S5b and reveal a thickness of ~100 nm both for Al and InAs layers. Figures 3b,c show Scanning Electron Microscopy (SEM) images of JJs with different widths and lengths, respectively. Also, in this case, straight edges of the MESA are appreciable, along with a clean separation of the JJ from Al wet etching residuals.

Samples were measured in a dilution fridge equipped with a z-axis superconducting magnet and a DC measurement setup; the electrical characterization was performed at 50 mK unless stated otherwise. Figure 3d shows forward and backward V-I characteristics of a JJ with L = 350 nm, W = 10 μm, and $n_{2D}$ ~ $10^{13}$ $cm^{-2}$, while V-I curves for JJ with different lengths, widths, and InAs sheet electron densities are reported in Figure S6. The JJ features an odd V-I curve and, at the switching current ($I_S$), switches from the dissipationless to the dissipative regime, exhibiting a normal-state resistance ($R_N = \frac{dV}{dI}|_{I>I_S}$). This demonstrates the effective proximization of the InAs layer and confirms the formation of the desired super-semi-super JJ. Once dissipative, the JJ returns to the dissipationless state when the bias current is lowered below the re-trapping current ($I_R \leq I_S$). It is worth noting that both $I_S$ and $I_R$ feature the same values for positive and negative bias currents ($I_S$ = $I_S^+$ = $I_S^-$ and $I_R$ = $I_R^+$ = $I_R^-$). The switching current value increases linearly with the JJ width, regardless of the length and the InAs sheet electron density, allowing the definition of the switching current density ($I_S$/W) as the switching current divided by the JJ width (Figure 3e). $I_S$/W exponentially decreases with the JJ length, as reported in Figure 3f top, where a monotonic suppression of the critical current density from 3.4 μA/μm with L = 350 nm to 1.27 μA/μm with L = 900 nm was observed in the case of $n_{2D}$ ~ $10^{13}$ $cm^{-2}$. Similar trends were also observed for InAs samples with different sheet electron densities. The $I_S$/W vs. L trend indicates that all the JJs featuring L ≥ 350 nm are not in the short junction regime with $L_{eff} \ll \xi_N$, where $L_{eff}$ is the JJ effective length, and $\xi_N$ is the InAs coherence length [65]. At 50 mK, we estimated a $\xi_N$ value of about 1.0 μm, 1.7 μm, and 2.2 μm for $n_{2D}$ ~ $10^{12}$, $10^{13}$, and $10^{14}$ $cm^{-2}$, respectively. These calculations, together with the experimental results, indicate that the JJ effective length is significantly longer than the inter-electrode separation, which can be explained by supercurrent contributions of source-to-drain paths longer than the inter-electrode separation. This stems from the superconducting proximization of the InAs epilayer underneath the Al leads, resulting in a junction not strictly defined by the inter-electrode separation [66]. Figure 3f bottom shows the relationship between $I_S/W$ and the InAs sheet electron density for a JJ length of 350 nm. The critical current density logarithmically increases from 0.5 μA/μm for $n_{2D} \sim 10^{12}$ cm$^{-2}$ to 7.3 μA/μm for $n_{2D} \sim 10^{14}$ cm$^{-2}$. The latter value is significantly higher than what is reported for JJs obtained via the Al-proximity effect on InAs 2D QWs and InAs 1D NWs, approaching levels typically achieved with InAs 3D substrates employing Nb as a superconductor (Figure 1a, red stars, Table S1). Furthermore, $I_S/W$ can be easily tuned by orders of magnitude, changing both the JJ morphological properties and the sheet electron density in the InAs epilayer. Figure S7a shows normal state resistances of JJs with different $n_{2D}$ and W, while L ~ 350 nm. A hyperbolic reduction of $R_N$ was observed with an increase in the JJ width regardless of the InAs sheet electron density, which agrees with the relationship $R_N = \rho \frac{L}{W \times t}$, where $t$ is the InAs epilayer thickness. $R_N$ increases by about ten times, decreasing $n_{2D}$ from $10^{14}$ cm$^{-2}$ to $10^{12}$ cm$^{-2}$, a trend consistent with the resistivity increment reported in Figure 2e (minor deviations could be related to physical lengths different from 350 nm). Based on the switching currents and normal state resistances achieved, the JJs show a critical voltage ($V_C = I_S \times R_N$) ranging from 50 to 80 μV, regardless of length, width, and InAs sheet electron density (Figure S7b). In-future improvements of InAsOI-based JJs, including L reduction to reach the short junction regime accomplished by using superconducting leads with higher critical temperatures, could increase the critical voltage reaching the mV range as reported for other InAs-based platforms [39][42].

We now discuss the temperature-dependent behavior of the JJs. Figure 3g shows forward and backward V-I characteristics of a JJ with L = 350 nm, W = 10 μm, and $n_{2D} \sim 10^{13}$ cm$^{-2}$ in the temperature range from 0.05 K to 1.5 K. As expected, the switching current and current density monotonically decreases with the increase in temperature, regardless of the JJ length (Figure 3h) and InAs sheet electron density (Figure 3i). Specifically, a strong suppression of $I_S$ with temperature was observed increasing the JJ length and decreasing the InAs sheet electron density. Additional insights on the $I_S$ vs. T curves can be found in the Supporting Information. In the case of $n_{2D} \sim 10^{13}$ cm$^{-2}$ (Figure 3h), JJs with L > 500 nm exhibit a fully dissipative behavior for temperatures > 1 K, while supercurrent was observed for L ≤ 500 nm up to 1.25 K. On the other hand, with the most minor inter-electrodes separation tested in this work, namely 350 nm, JJs fabricated with $n_{2D} \geq 10^{13}$ cm$^{-2}$ feature $I_S$ up to 1.25 K, while JJs manufactured with $n_{2D} \sim 10^{12}$ cm$^{-2}$ exhibit a fully-dissipative behavior for temperatures > 1 K (Figure 3i). The latter result agrees with the resistance vs. temperature behavior shown in Figure S8. JJs with L = 350 nm fabricated on InAsOIs with $n_{2D} \geq 10^{13}$ cm$^{-2}$ show critical temperature ($T_C$) equal to that of Al, i.e., 1.22 K. On the other hand, the JJs manufactured with the $10^{12}$ $cm^{-2}$ sample feature $T_C$ = 0.93 K. The obtained critical temperatures suggest that the usage of a superconductor with a higher $T_C$ (e.g., Nb) can allow to operate with InAsOI at liquid He temperatures removing the need of a dilution cryostat. We also evaluated the JJ thermal inertia [67] by collecting the temperature-dependent behavior of the re-trapping current and assessing the percentage ratio $(I_S-I_R)/I_S$: the lower the ratio, the lower the JJ thermal inertia, and vice versa. Figure S9 shows $(I_S-I_R)/I_S$ calculated for JJs with different lengths, widths, and InAs sheet electron densities. At 50 mK, the JJs feature a 50 ÷ 75 % ratio, which progressively reduces to 0 % at 750 mK, regardless of the JJ length, width, and InAs sheet electron density.

We next evaluated the JJ behavior under different values of the out-of-plane magnetic field ($B_\perp$) since the modulation of the switching current by $B_\perp$-mediated quantum interference is an essential hallmark of the Josephson effect. Figures 4,S10 shows the switching current vs. $B_\perp$ trend for different JJs with different lengths, widths, InAs sheet electron densities, and temperatures.

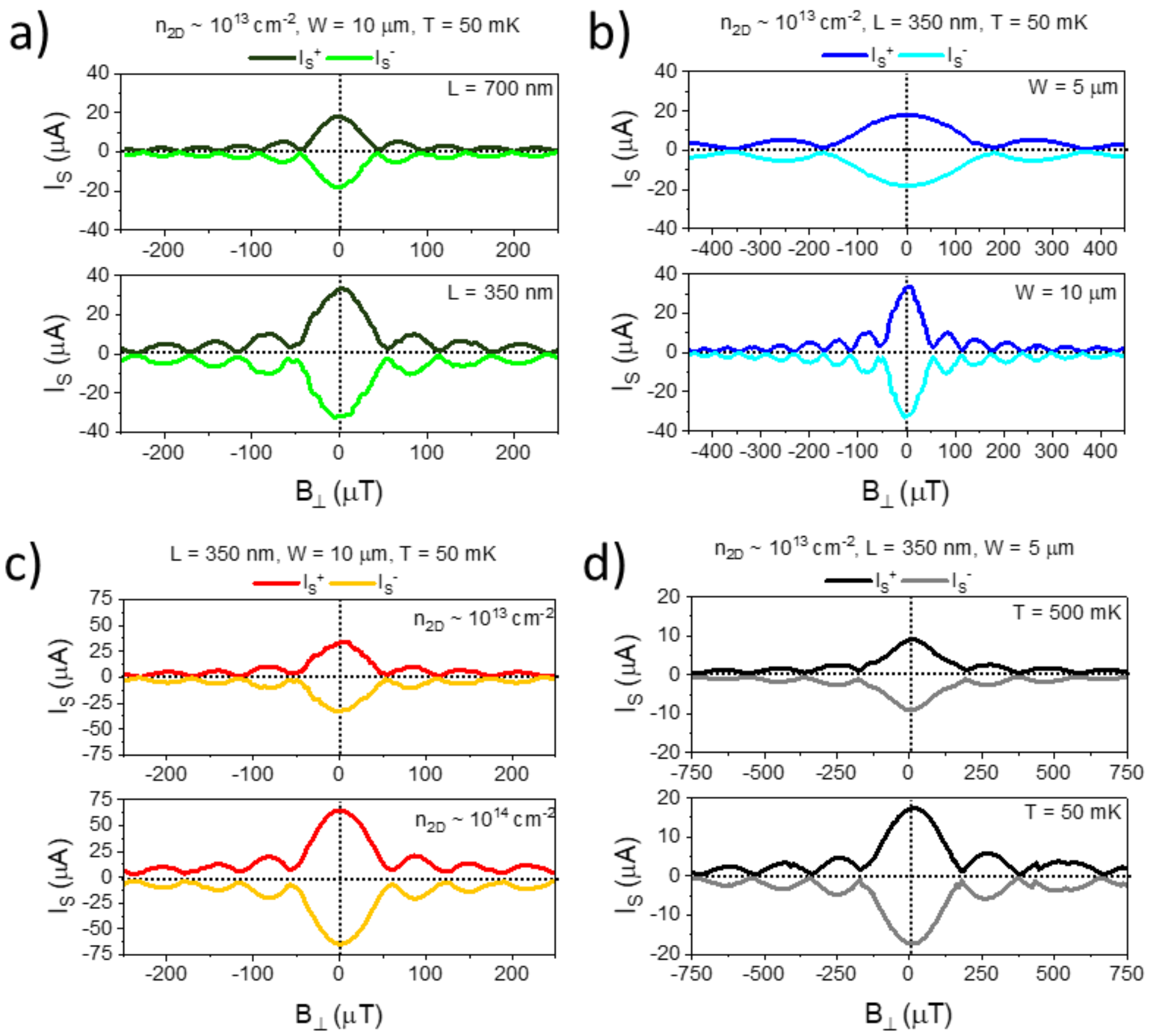

**Figure 4: InAsOI-based Josephson Junctions Fraunhofer diffraction patterns.** Switching current vs. out-of-plane magnetic field for JJs with different lengths (a), widths (b), InAs sheet electron densities (c), and temperatures (d). JJs specific properties are reported in each panel.

Regardless of morphology, InAs sheet electron density, and temperature, by applying $B_\perp$ to the JJ, the switching current follows a characteristic Fraunhofer-like pattern $I_S(B_\perp) = I_S|_{B_\perp=0T} \times \left|\frac{\sin(\pi B_\perp A_{eff}/\Phi_0)}{\pi B_\perp A_{eff}/\Phi_0}\right|$, where $A_{eff} = \gamma \times W \times L$ is the effective JJ area, and $\Phi_0 = {h}/{2e}$ is the flux quantum. $\gamma$ is a scaling factor which takes into account the magnetic field focusing and the enlargement of the magnetic effective JJ length compared to the inter-electrode separation. The sinc-like Fraunhofer pattern behavior is expected for a SNS junction with $W \gg \xi_0$, where $\xi_0$ is the InAs coherence length relevant for modeling the vortex state within the SNS JJ [68][69] . $\xi_0$ is estimated as 372 nm, 619 nm, and 800 nm for InAs epilayer with $n_{2D} \sim 10^{12}$ cm$^{-2}$, $10^{13}$ cm$^{-2}$, and $10^{14}$ cm$^{-2}$, respectively. This is also consistent with observations for InAs 2D QW JJs [17][43][70][41][71]. The magnitude of the interference pattern increases by reducing the JJ length (Figure 4a) and temperature (Figure 4d) and by increasing the JJ width (Figure 4b) and the InAs sheet electron density (Figure 4c). On the other hand, the interference pattern periodicity, which is related to $A_{eff}$, increases both enlarging the JJ length (Figure 4a) and width (Figure 4b). In our observations, we find that the periodicity of the Fraunhofer pattern deviates from the expected theoretical value of $B_\perp = \Phi_0/W \times L$. The observed periodicity, in relation to the applied magnetic field, is notably smaller than anticipated. As previously said, this discrepancy can be attributed to $\gamma > 1$ [72][73]. Figure S11 reports the calculation of $\gamma$ as a function of the JJ aspect ratio W/L. As expected [74], $\gamma$ linearly depends on W/L, with an estimated value of 1 (absence of magnetic field focusing) when W/L=0. We also notice a slight increase in the minima values of the Fraunhofer patterns near the central lobe, which is typical in diffusive SNS junctions where the switching current values of the minima follow a Gaussian envelope [75][76]. Eventually, we extended the out-of-plane magnetic field range of the Fraunhofer diffraction pattern shown in Figure 4a bottom to collect the interference minima (Figure S12). We defined the switching current suppression factor (SF%) as $SF\% = 100 \times \frac{I_{S\,MAX}-I_{S\,min}}{I_{S\,MAX}}$, and we found SF% = 99.55 %, which was obtained with a $B_\perp$ sweep of 960 μT.

## Conclusions

In this work, we proposed the InAs-on-insulator (InAsOI) as a groundbreaking platform to develop superconducting electronics. An epilayer of semiconducting InAs with different electron densities was grown onto an InAlAs metamorphic buffer layer, efficiently used as a cryogenic insulator to electrically decouple adjacent devices. Josephson Junctions with various lengths and widths were fabricated using Al as a superconductor and InAsOI with different electron densities. The switching current and supercurrent density can be easily tuned by orders of magnitude by changing both the JJ morphological properties and the InAs electron density. We achieved a switching current density of 7.3 μA/μm, a critical voltage of 50-to-80 μV, and a critical temperature equal to that of the superconductor used. The usage of superconductors with higher critical temperatures can be envisaged to employ InAsOI at liquid He temperatures. For all the JJs, the switching current follows a characteristic Fraunhofer pattern with the out-of-plane magnetic field, from which a switching current suppression factor of 99.95 % was calculated. These achievements open up the use of the InAsOI platform to design and fabricate surface-exposed Josephson Field Effect Transistors with high critical current densities and superior gating properties. Moreover, the electron density of the InAs epilayer can be locally tuned post-heterostructure growth via dopant implantation or thermal diffusion to obtain quantum devices with different properties on the same substrate.

## Experimental Section

### *InAsOI Heterostructure Growth via Molecular Beam Epitaxy*

InAsOI was grown on semi-insulating GaAs (100) substrates using solid-source MBE. Starting from the GaAs substrate, the sequence of the layer structure includes a 200 nm-thick GaAs layer, a 200 nm-thick GaAs/$Al_{0.16}Ga_{0.84}As$ superlattice, a 200 nm-thick GaAs layer, a 1.250 μm-thick step-graded $In_XAl_{1-X}As$ metamorphic buffer (with X increasing from 0.15 to 0.81), a 400 nm-thick $In_{0.84}Al_{0.16}As$ overshoot layer, and a 100 nm-thick InAs layer. The metamorphic buffer and the overshoot layers were grown at optimized substrate temperatures of 320 °C ± 5 °C. The InAs epilayer was grown at 480 ± 5 °C. The doping of the InAs layer is achieved by using Si cell at different temperatures, namely 1230 and 1300 °C.

### *InAsOI Josephson Junction Fabrication*

Al-InAs-Al JJs with several widths (W = 5, 10, 20 μm) and inter-electrode separations (length, L = 350, 500, 700, 900 nm) where fabricated using InAsOIs with different sheet electron densities. First, InAsOI substrates were etched from native InAs oxide ($InAsO_X$) and passivated with S-termination dipping the InAsOI samples in a $(NH_4)_2S_X$ solution (290 mM $(NH_4)_2S$ and 300 mM S in DIW). Then, a 100-nm-thick Al layer was deposited at a rate of 2 A/s at a residual chamber pressure of 1E-8 ÷ 5E-9 Torr. JJs were fabricated via two aligned lithographic steps: first, Al and InAs MESA were defined by UV-lithography and manufactured by successive Al and InAs wet etching, setting the JJ width and leaving the cryogenic-dielectric-InAlAs layer exposed. Al was removed by dipping samples in the Transene Al Etchant Type D, while InAs was etched by dipping samples in a $H_3PO_4$:$H_2O_2$ solution (348 mM $H_3PO_4$, 305 mM $H_2O_2$ in DIW). The JJ length was then defined by electron-beam-lithography and Al wet-etching, leaving the underneath InAs unaffected.

**Supporting Information**

Supporting Information is available from the Wiley Online Library.

**Acknowledgments**

We thank Stefan Heun for helping to carry out room-temperature and cryogenic measurements to estimate InAs resistivity, mobility, and two-dimensional charge density. This work was supported in part by EU's Horizon 2020 Research and Innovation Framework Program under Grant 964398 (SUPERGATE), by Grant 101057977 (SPECTRUM), and in part by the Piano Nazionale di Ripresa e Resilienza, Ministero dell'Università e della Ricerca (PNRR MUR) Project under Grant PE0000023-NQSTI.

**Supporting Information**

# InAs on Insulator: A New Platform for Cryogenic Hybrid Superconducting Electronics

*Alessandro Paghi**, *Giacomo Trupiano, Giorgio De Simoni, Omer Arif, Lucia Sorba, and Francesco Giazotto*

Istituto Nanoscienze-CNR and Scuola Normale Superiore, Piazza San Silvestro 12, 56127 Pisa, Italy.

*Corresponding authors: alessandro.paghi@nano.cnr.it

**Summary**

## 1. Supporting Tables

**Table S1: State-of-art of the InAs-based superconductive platforms.** For each platform, namely InAsOIs, InAs 3D substrates, InAs 2D QWs, and InAs 1D NWs, the most representative JJs are reported. The best devices for each category have red character color and *italic* format. In the case of InAs 1D NWs, “W” stays for the NW diameter.

| InAs Specs | Superconductor | L [nm] | W [µm] | Ic [µA] | Ic/W [µA/µm] | Vc [µV] | T [mK] | Ref. |
|---|---|---|---|---|---|---|---|---|
| **InAsOI** | | | | | | | | |
| *t = 100 nn, $\mu_n$ = 3190 cm$^2$/Vs $n_{2D}$ = 1.4E14 cm$^{-2}$* | *Al* | *350* | *19.9* | *139* | *7.0* | *82* | *50* | *This work* |
| t = 100 nn, $\mu_n$ = 3190 cm$^2$/Vs $n_{2D}$ = 1.4E14 cm$^{-2}$ | Al | 500 | 19.8 | 84 | 4.2 | 51 | 50 | This work |
| t = 100 nn, $\mu_n$ = 3190 cm$^2$/Vs $n_{2D}$ = 1.4E14 cm$^{-2}$ | Al | 700 | 19.9 | 66 | 3.3 | 47 | 50 | This work |
| t = 100 nn, $\mu_n$ = 3190 cm$^2$/Vs $n_{2D}$ = 1.4E14 cm$^{-2}$ | Al | 900 | 19.9 | 59 | 3.0 | 61 | 50 | This work |
| t = 100 nn, $\mu_n$ = 6870 cm$^2$/Vs $n_{2D}$ = 2.0E13 cm$^{-2}$ | Al | 350 | 19.5 | 66 | 3.4 | 71 | 50 | This work |
| t = 100 nn, $\mu_n$ = 6870 cm$^2$/Vs $n_{2D}$ = 2.0E13 cm$^{-2}$ | Al | 500 | 19.2 | 52 | 2.7 | 72 | 50 | This work |
| t = 100 nn, $\mu_n$ = 6870 cm$^2$/Vs $n_{2D}$ = 2.0E13 cm$^{-2}$ | Al | 700 | 19.8 | 34 | 1.7 | 56 | 50 | This work |
| t = 100 nn, $\mu_n$ = 6870 cm$^2$/Vs $n_{2D}$ = 2.0E13 cm$^{-2}$ | Al | 900 | 19.5 | 25.5 | 1.3 | 62 | 50 | This work |
| t = 100 nn, $\mu_n$ = 7863 cm$^2$/Vs $n_{2D}$ = 3.8E12 cm$^{-2}$ | Al | 350 | 21.2 | 8.2 | 0.39 | 57 | 50 | This work |
| t = 100 nn, $\mu_n$ = 7863 cm$^2$/Vs $n_{2D}$ = 3.8E12 cm$^{-2}$ | Al | 500 | 20.8 | 1.5 | 0.07 | 20 | 50 | This work |
| t = 100 nn, $\mu_n$ = 7863 cm$^2$/Vs $n_{2D}$ = 3.8E12 cm$^{-2}$ | Al | 700 | 21.8 | 0.7 | 0.03 | 12 | 50 | This work |
| **InAs 3D Substrates** | | | | | | | | |
| n-type, $n_{3D}$ = 2.6E18 cm$^{-3}$ | Nb | 380 | 80 | 700 | 8.75 | 1.6 | 2000 | [1] |
| n-type, $n_{3D}$ = 2.6E18 cm$^{-3}$ | Nb | 420 | 80 | 500 | 6.25 | 1.2 | 2000 | [1] |
| n-type, $n_{3D}$ = 2.6E18 cm$^{-3}$ | Nb | 480 | 80 | 250 | 3.125 | 0.58 | 2000 | [1] |
| n-type, $n_{3D}$ = 2.6E18 cm$^{-3}$ | Nb | 540 | 80 | 100 | 1.25 | 0.23 | 2000 | [1] |
| n-type, $n_{3D}$ = 2.6E18 cm$^{-3}$ | Nb | 560 | 80 | 180 | 2.25 | 0.41 | 2000 | [1] |
| n-type, $n_{3D}$ = 2.6E18 cm$^{-3}$ | Nb | 700 | 80 | 80 | 1 | 0.18 | 2000 | [1] |
| n-type, $n_{3D}$ = 2.6E18 cm$^{-3}$ | Nb | 825 | 80 | 40 | 0.5 | 0.09 | 2000 | [1] |
| n-type, $n_{3D}$ = 2.5E17 cm$^{-3}$ | Nb | 500 | 80 | 50 | 0.625 | 0.12 | 2000 | [1] |
| n-type, $n_{3D}$ = 2.5E17 cm$^{-3}$ | Nb | 700 | 80 | 10 | 0.125 | 0.02 | 2000 | [1] |
| n-type, $n_{3D}$ = 2.4E16 cm$^{-3}$ | Nb | 500 | 80 | 5.5 | 0.06875 | 0.01 | 2000 | [1] |
| n-type, $n_{3D}$ = 2.4E16 cm$^{-3}$ | Nb | 650 | 80 | 2 | 0.025 | 0.005 | 2000 | [1] |
| *p-type* | *Nb* | *20* | *43* | *870* | *20.23* | *1350* | *2000* | *[2]* |
| p-type | Nb | 60 | 43 | 600 | 1.95 | 1100 | 2000 | [2] |
| p-type | Nb | 40 | 43 | 100 | 2.33 | 700 | 2000 | [2] |
| p-type, $h_{3D}$ = 2.6E15 $cm^{-3}$ | Nb | 400 | 80 | 0.9 | 0.011 | 63 | 20 | [3] |
| | **InAs 2D QWs** | | | | | | | |
| t = 8 nm, $\mu_n$ = 18000 $cm^2/Vs$ $n_{2D}$ = 8.0E11 $cm^{-2}$ | Al | 80 | 2.5 | 1 | 0.4 | 200 | 10 | [4] |
| t = 7 nm, $\mu_n$ = 12000 $cm^2/Vs$ $n_{2D}$ = 1.6E12 $cm^{-2}$ | Al | 150 | 2.5 | 1 | 0.4 | 15.5 | 20 | [5] |
| t = 4 nm, $\mu_n$ = 14400 $cm^2/Vs$ | Al | 100 | 4 | 5 | 1.25 | 486 | 20 | [6] |
| - | Al | 150 | 4 | 1.4 | 0.35 | - | 30 | [7] |
| t = 5 nm, $\mu_n$ = 25000 $cm^2/Vs$ $n_{2D}$ = 1.0E12 $cm^{-2}$ | Al | 140 | 5 | 1.5 | 0.3 | 400 | 50 | [8][9] |
| t = 7 nm, $\mu_n$ = 130000 $cm^2/Vs$ $n_{2D}$ = 5.0E11 $cm^{-2}$ | Al | 300 | 9 | 1.1 | 0.12 | 30 | 20 | [10] |
| - | Al | 50 | 5 | 0.75 | 0.15 | - | 20 | [11] |
| t = 7 nm, $\mu_n$ = 16800 $cm^2/Vs$ $n_{2D}$ = 9.6E11 $cm^{-2}$ | Al | 150 | 4 | 1.8 | 0.45 | 160 | 30 | [12] |
| - | Al | 100 | 3.15 | 2.4 | 0.76 | 295 | 100 | [13] |
| t = 7 nm, $\mu_n$ = 15000 $cm^2/Vs$ $n_{2D}$ = 1.2E12 $cm^{-2}$ | Al | 150 | 2 | 0.01 | 0.005 | 3 | 55 | [14] |
| t = 4 nm (hbN gate insulator) $n_{2D}$ = 7.0E11 $cm^{-2}$ | Al | 100 | 3 | 2.7 | 0.9 | 350 | 30 | [15] |
| *t = 4 nm ($Al_2O_3$ gate insulator) $n_{2D}$ = 7.0E11 $cm^{-2}$* | *Al* | *100* | *5* | *2.7* | *1.67* | *-* | *30* | *[15]* |
| $\mu_n$ = 14500 $cm^2/Vs$ $n_{2D}$ = 1.0E12 $cm^{-2}$ | Al | 100 | 35 | 2 | 0.057 | - | 30 | [16] |
| t = 7 nm, $\mu_n$ = 17500 $cm^2/Vs$ $n_{2D}$ = 1.0E12 $cm^{-2}$ | Al | 200 | 3 | 1.4 | 0.47 | 135 | 30 | [17] |
| t = 30 nm, $\mu_n$ = 3200 $cm^2/Vs$ | Al | 120 | 0.32 | 0.075 | 0.234 | 83 | 30 | [18] |
| t = 7 nm, $\mu_n$ = 15600 $cm^2/Vs$ $n_{2D}$ = 1.26E12 $cm^{-2}$ | Al | 250 | 3 | 1.77 | 0.59 | 165 | 30 | [19] |
| t = 4 nm, $\mu_n$ = 111000 $cm^2/Vs$ $n_{2D}$ = 2.3E12 $cm^{-2}$ | Nb | 350 | 40 | 5 | 0.125 | 80 | 1000 | [20] |
| - | Nb | 1200 | 100 | 75 | 0.75 | 32 | 1400 | [21] |
| t = 15 nm, $\mu_n$ = 100000 $cm^2/Vs$ $n_{2D}$ = 2.1E12 $cm^{-2}$ | Nb | 470 | 0.7 | 0.26 | 0.37 | 78 | 100 | [22] |
| t = 15 nm | Nb | 600 | 0.5 | 0.6 | 1.2 | 400 | - | [23] |
| t =4 nm, $\mu_n$ = 160000 $cm^2/Vs$ $n_{2D}$ = 6.24E11 $cm^{-2}$ | Nb | 1100 | 0.85 | 0.15 | 0.18 | 140 | 400 | [24] |
| *t =9 nm, $\mu_n$ = 10000 $cm^2/Vs$ $n_{2D}$ = 1E12 $cm^{-2}$* | *Al-NbTi* | *150* | *4.3* | *7.5* | *1.75* | *400* | *35* | *[25]* |
| t =4 nm, $\mu_n$ = 155000 $cm^2/Vs$ $n_{2D}$ = 1.86E12 $cm^{-2}$ | Nb | 400 | 80 | 21 | 0.27 | 52.5 | 1000 | [26] |
| | **InAs 1D NWs** | | | | | | | |
| *-* | *Al* | *30* | *0.08* | *0.8* | *10* | *128* | *15* | *[27]* |
| - | Al | 90 | 0.08 | 0.095 | 1.1875 | 54 | 15 | [27] |
| - | Al | 100 | 0.08 | 0.054 | 0.675 | 30 | 15 | [27] |
| - | Al | 220 | 0.08 | 0.03 | 0.375 | 31 | 15 | [27] |
| - | Al | 500 | 0.08 | 0.003 | 0.0375 | 12.6 | 15 | [27] |
| - | Al | 600 | 0.08 | 0.001 | 0.0125 | 4.8 | 15 | [27] |
| - | Al | 144 | 0.08 | 0.0085 | 0.11 | - | 20 | [28] |
| - | Al | 80 | 0.05 | 0.01 | 0.2 | 70 | 15 | [29] |
| - | Al | 250 | 0.1 | 0.13 | 1.3 | 43 | 30 | [30] |
| $\mu_n = 2000\ cm^2/Vs$<br>$n_{3D} = 1.0E19\ cm^{-3}$ | Al | 100 | 0.07 | 0.1 | 1.43 | 60 | 40 | [31] |
| - | Sn | 142 | 0.176 | 0.3 | 1.71 | 100 | 40 | [32] |
| *$\mu_n = 2000\ cm^2/Vs$*<br>*$n_{3D} = 3.0E18\ cm^{-3}$* | *Ti/Pb* | *100* | *0.08* | *0.615* | *7.69* | *250* | *10* | *[33]* |
| $n_{3D} = 7.0E18\ cm^{-3}$ | Nb | 140 | 0.11 | 0.1 | 0.91 | 75 | 400 | [34] |
| $\mu_n = 970\ cm^2/Vs$<br>$n_{3D} = 1.0E18\ cm^{-3}$ | Nb | 70 | 0.08 | 0.0028 | 0.035 | 10.6 | 400 | [34] |
| - | Nb | 55 | 0.08 | 0.075 | 0.94 | 214 | 15 | [35] |
| - | Ta | 280 | 0.065 | 0.002 | 0.031 | 3.2 | 15 | [36] |

## 2. Supporting Figures

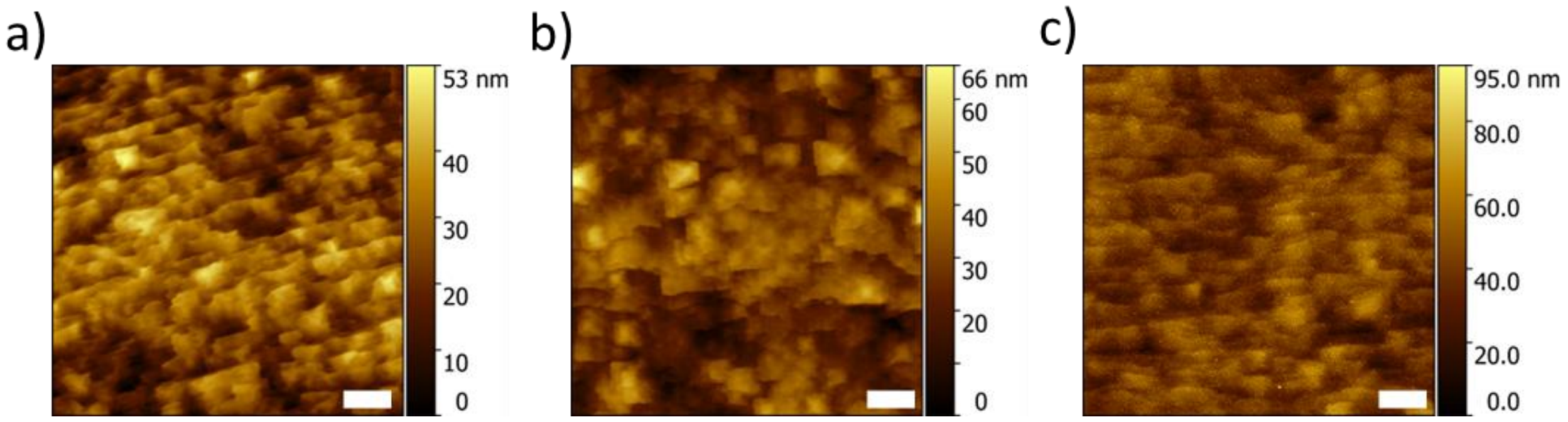

**Figure S1: Atomic force microscopy photographs of InAsOIs.** a,b) InAsOI featuring an InAs epilayer with $n_{2D}$ ~ $10^{12}$ $cm^{-2}$ (a) and $n_{2D}$ ~ $10^{14}$ $cm^{-2}$ (b). c) InAsOI featuring an InAs epilayer with $n_{2D}$ ~ $10^{13}$ $cm^{-2}$ and a 100 nm-thick Al film. The scalebar is 2.5 μm.

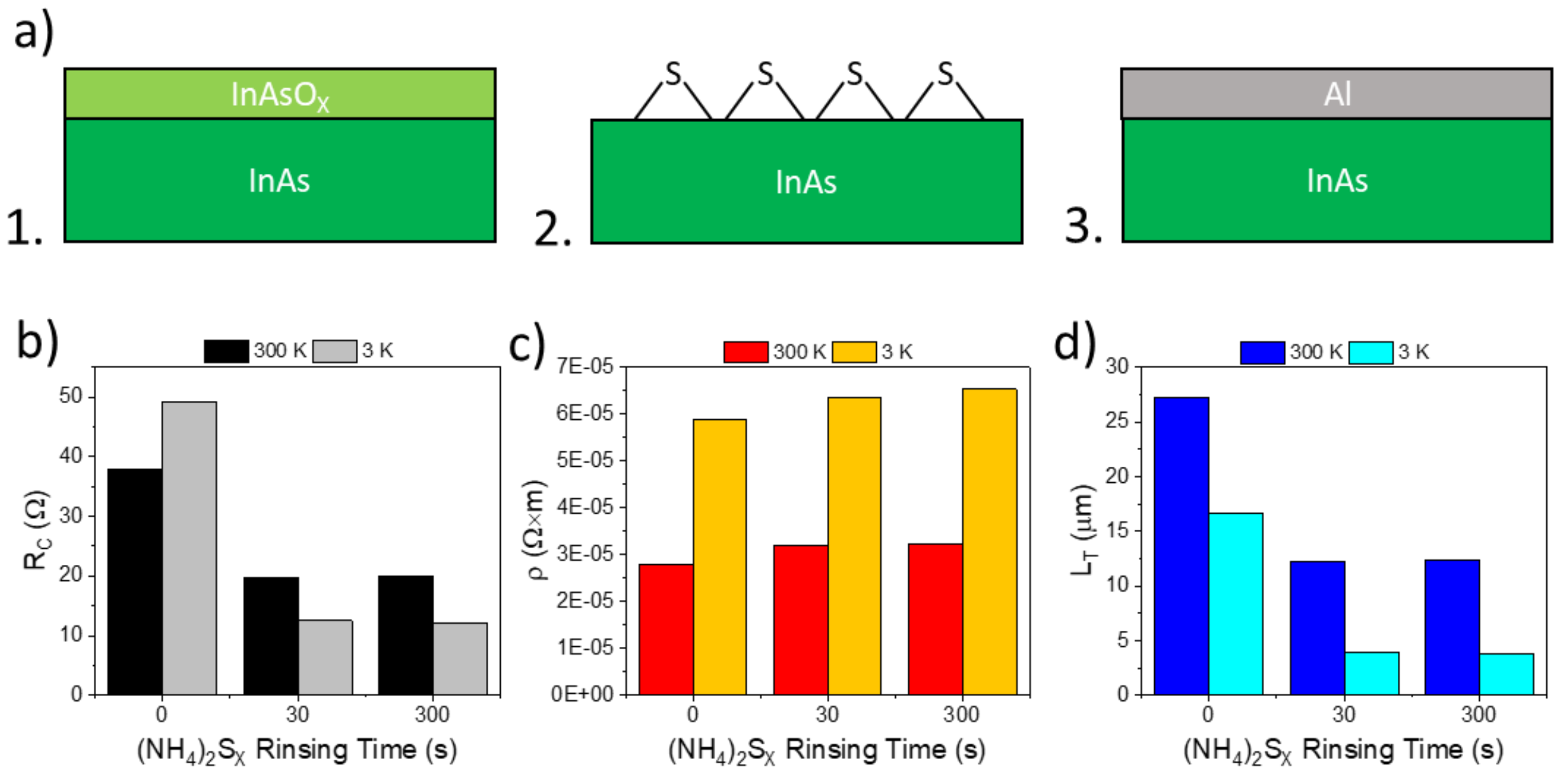

**Figure S2: Impact of the InAs native oxide ($InAsO_X$) removal and concomitant surface Sulfur passivation on the electrical properties of InAsOIs.** a) Schematic representation of the chemical etching and passivation process. (1) The surface exposed InAs epilayer features a native oxide layer ($InAsO_X$) due to the interaction with air. (2) The InAs surface was etched from $InAsO_X$ and passivated with S-termination, rinsing the sample with a $(NH_4)_2S_X$ solution. (3) An Al film (100 nm-thick) was immediately evaporated on the etched and passivated InAsOI. b) Al/InAsOI ($n_{2D}$ ~ $10^{12}$ $cm^{-2}$) contact resistance changing the $(NH_4)_2S_X$ solution rinsing time both at 300 K and 3K. The contact resistance at 3K decreases by 75%, dipping the sample in the $(NH_4)_2S_X$ solution, regardless of the rinsing time. c) InAsOI ($n_{2D}$ ~ $10^{12}$ $cm^{-2}$) resistivity changing the $(NH_4)_2S_X$ solution rinsing time both at 300 K and 3K. The resistivity is practically unaffected by the $(NH_4)_2S_X$ solution rinsing time. A little increase in the InAsOI resistivity could be due to a progressive removal of the epilayer thickness. d) Al/InAsOI ($n_{2D}$ ~ $10^{12}$ $cm^{-2}$) transmission length changing the $(NH_4)_2S_X$ solution rinsing time both at 300 K and 3K. The transmission length at 3K decreases by 75%, dipping the sample in the $(NH_4)_2S_X$ solution, regardless of the rinsing time.

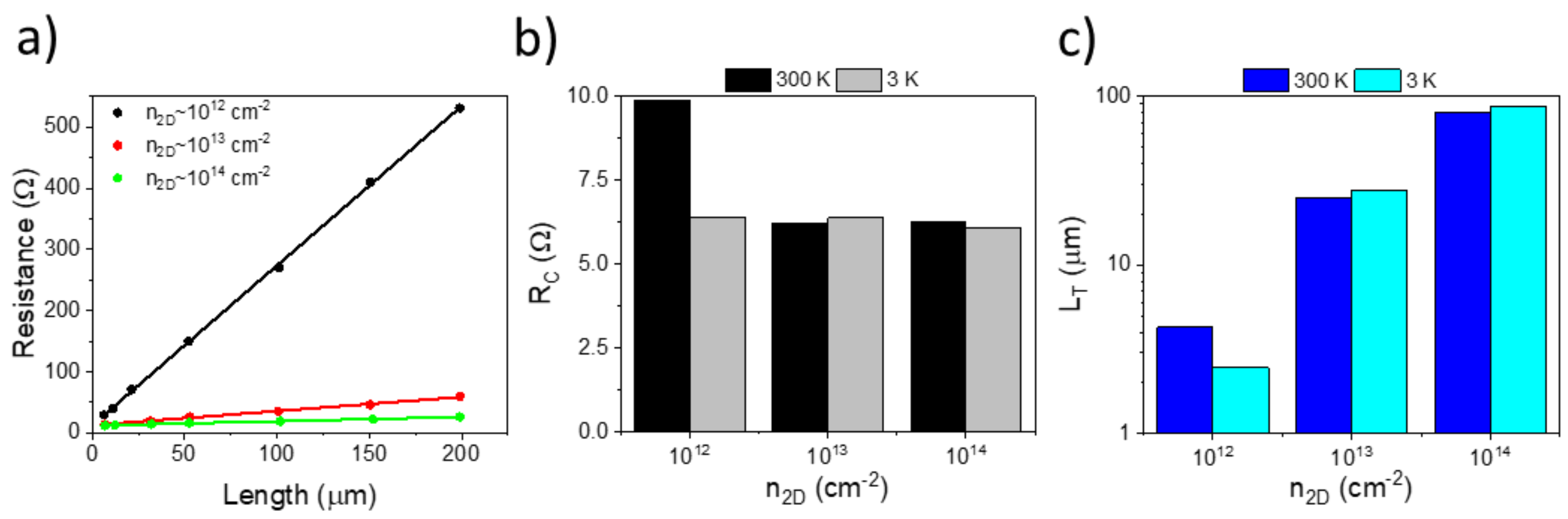

**Figure S3: Impact of the InAs native oxide (InAsOx) removal and concomitant surface Sulfur passivation on the electrical properties of InAsOIs with different sheet electron densities.** a) Resistance vs. TLM inter-pads lengths. The slope of the curve, which is related to the InAsOI resistivity, decreases in agreement with an increase in the InAsOI sheet electron density. b,c) Al/InAs contact resistance(b) and transmission length (c) for InAsOIs with different sheet electron densities both at 300 K and 3K. The contact resistance at 3K is practically unaffected by the InAsOI sheet electron density. The transmission length linearly increases with the InAsOI sheet electron density.

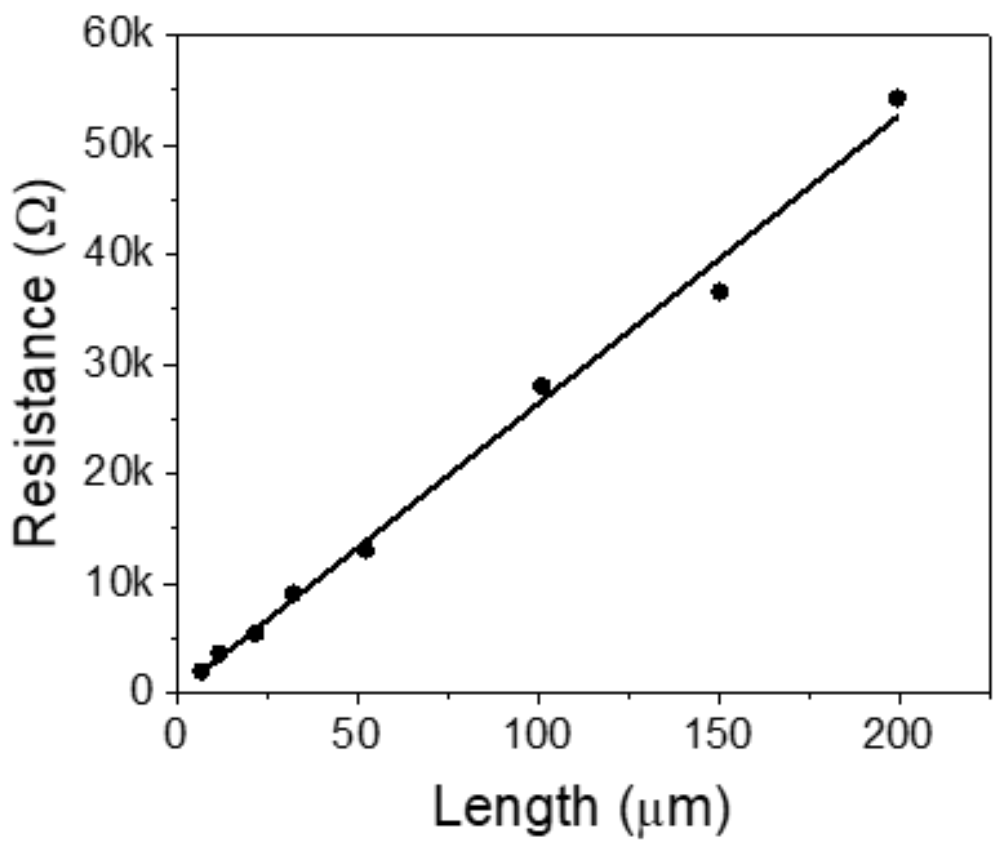

**Figure S4: Resistance vs. TLM inter-pads lengths for the InAlAs metamorphic buffer layer at 300 K.** The linear trend between resistance and length indicates that the InAlAs metamorphic buffer layer at room temperature features an ohmic behavior.

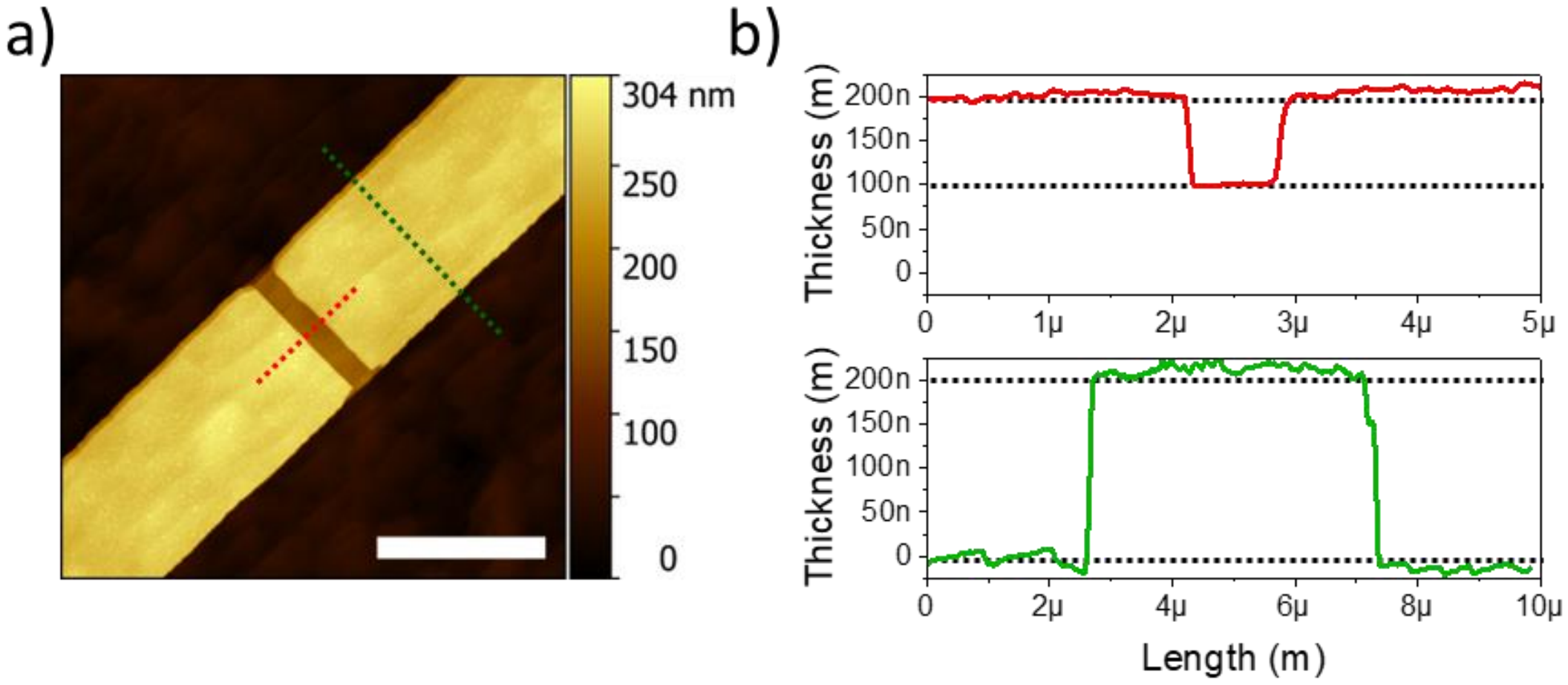

**Figure S5: Atomic force microscopy analysis of a Josephson Junction featuring W = 5 μm and L = 900 nm.** a) Atomic force microscopy photograph of the JJ. b). The thickness profile was extrapolated from the red dashed line in (a). Al film thickness of 100 nm is appreciated. c). The thickness profile was extrapolated from the green dashed line in (a). Al/InAs MESA thickness of 200 nm is appreciated. The scalebar is 5 μm.

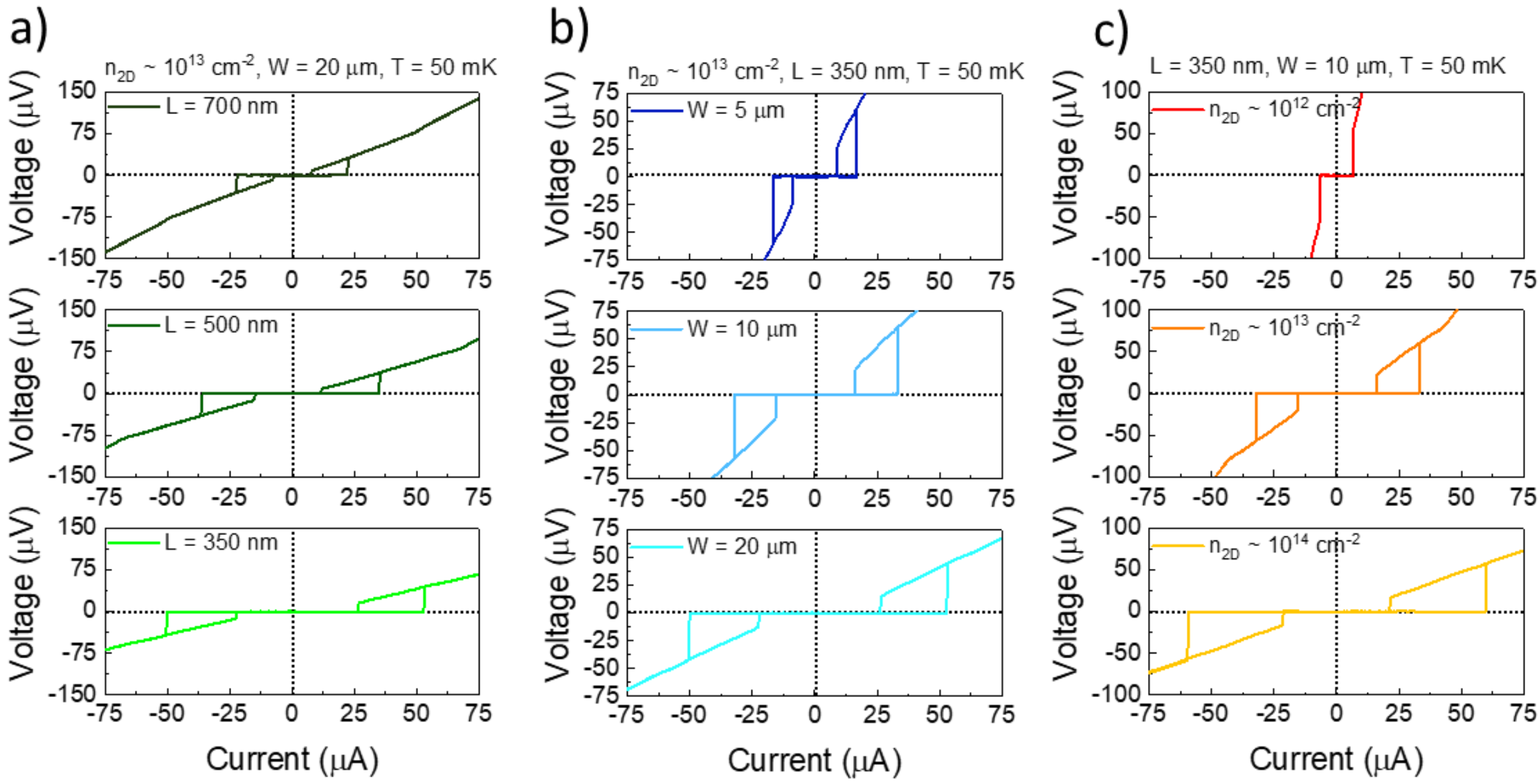

**Figure S6: Current vs. voltage curves of the fabricated Josephson Junctions at 50 mK.** a,b,c) I-V curves for JJs with different lengths (a), widths (b), and InAsOI sheet electron densities (c).

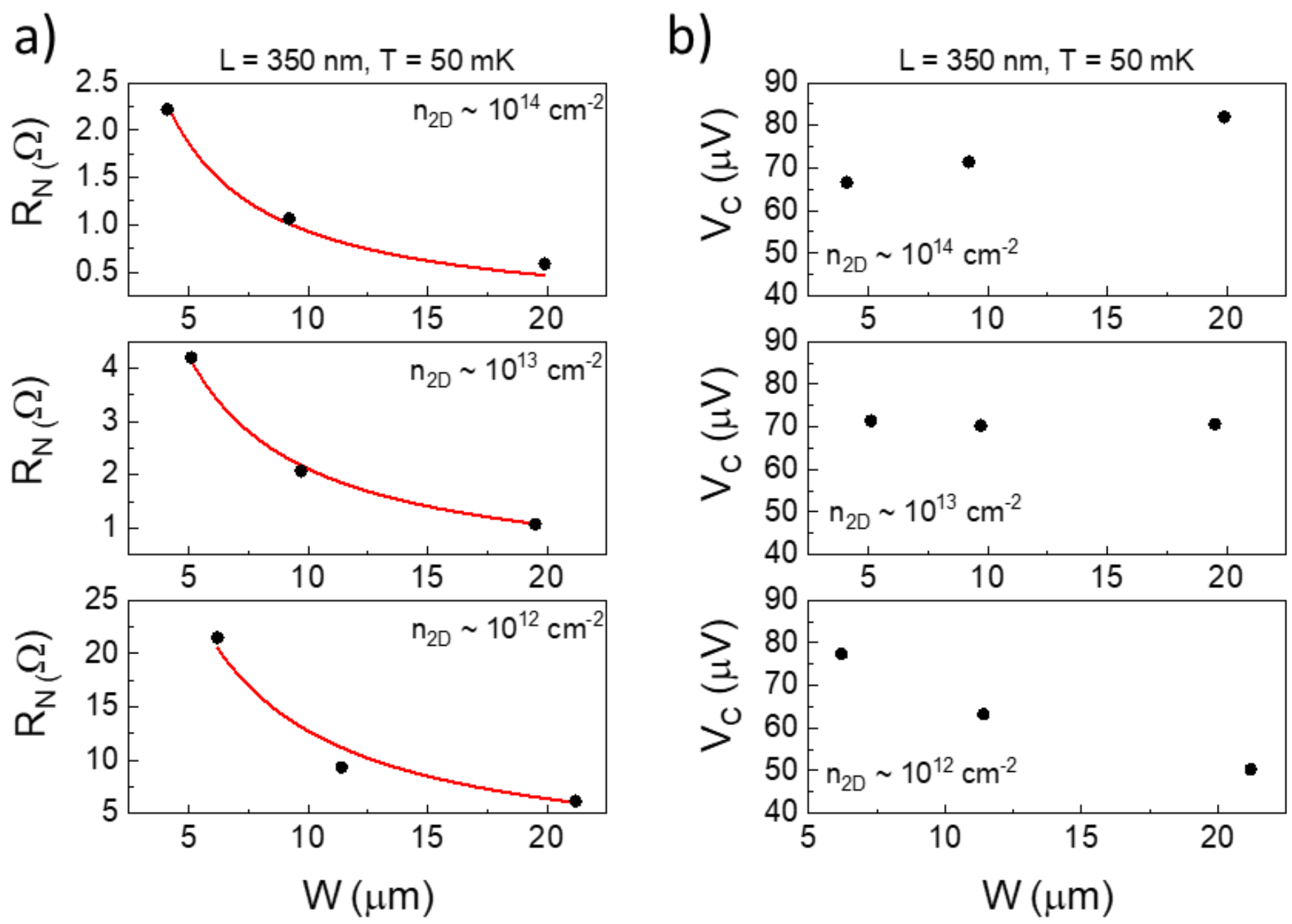

**Figure S7: Normal state resistance and critical voltage of the fabricated Josephson Junctions at 50 mK.** a) Normal state resistance vs. JJ width of InAsOIs with different sheet electron densities. As expected, a hyperbolic reduction of the normal state resistance was observed with an increase in the width, regardless of the InAsOI sheet electron density. b) Critical voltage vs. JJ width of InAsOIs with different sheet electron densities. The critical voltage ranges from 50 to 80 μV, regardless of the width and InAsOI electron density.

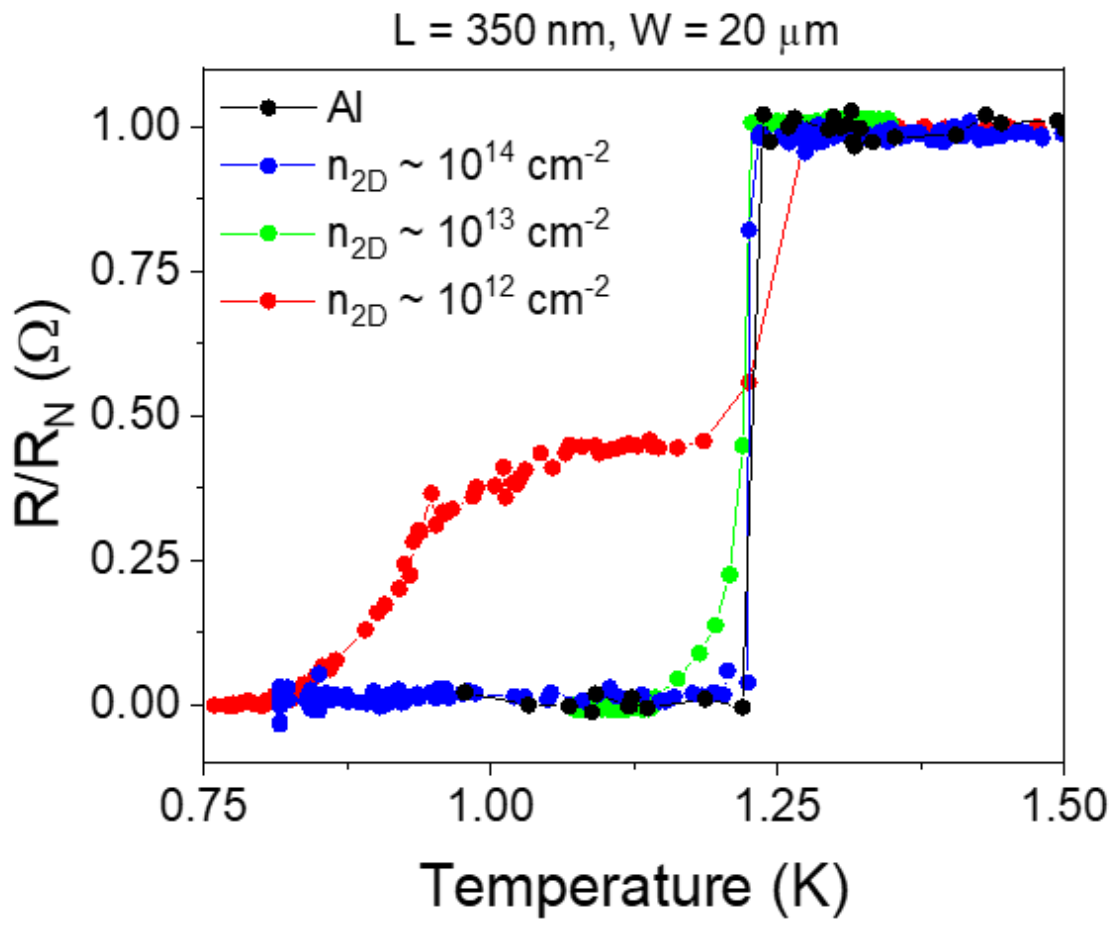

**Figure S8: Josephson Junctions (L=350 nm) resistances divided by the normal state resistances as a function of the temperature.** JJs fabricated on InAsOIs with $n_{2D} \geq 10^{13}$ $cm^{-2}$ show a critical temperature equal to that of Al, i.e., 1.22 K. On the other hand, JJs manufactured on InAsOIs with $n_{2D} \sim 10^{12}$ $cm^{-2}$ feature $T_C = 0.93$ K.

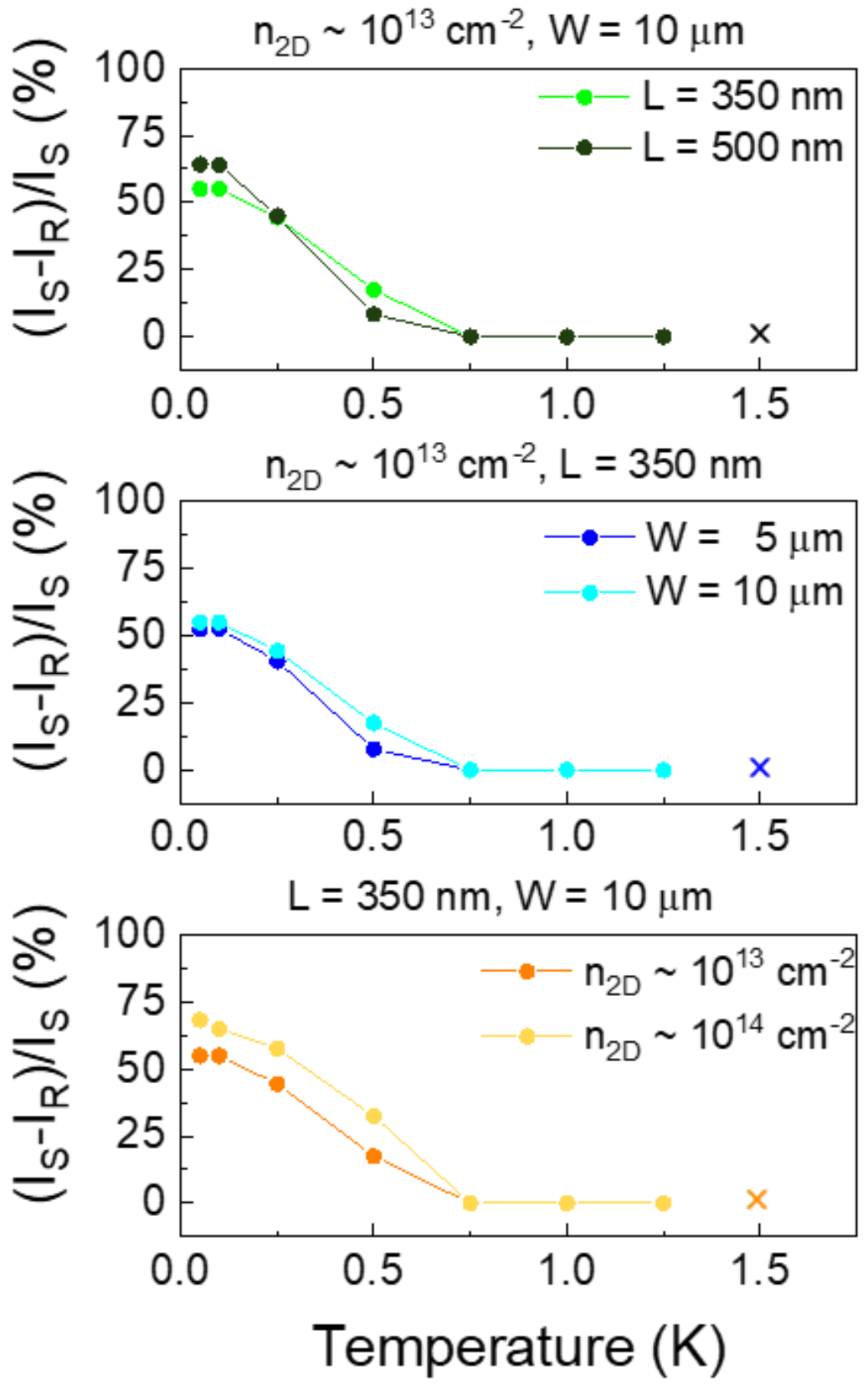

**Figure S9: Normalized switching - re-trapping current variation as a function of the temperature for Josephson Junctions with different lengths (top), widths (middle), and InAsOI sheet electron densities (bottom).** At 50 mK, the JJs feature a 50 ÷ 75 % ratio, which progressively reduces to 0 % at 750 mK, regardless of the JJ length, width, and InAsOI sheet electron density.

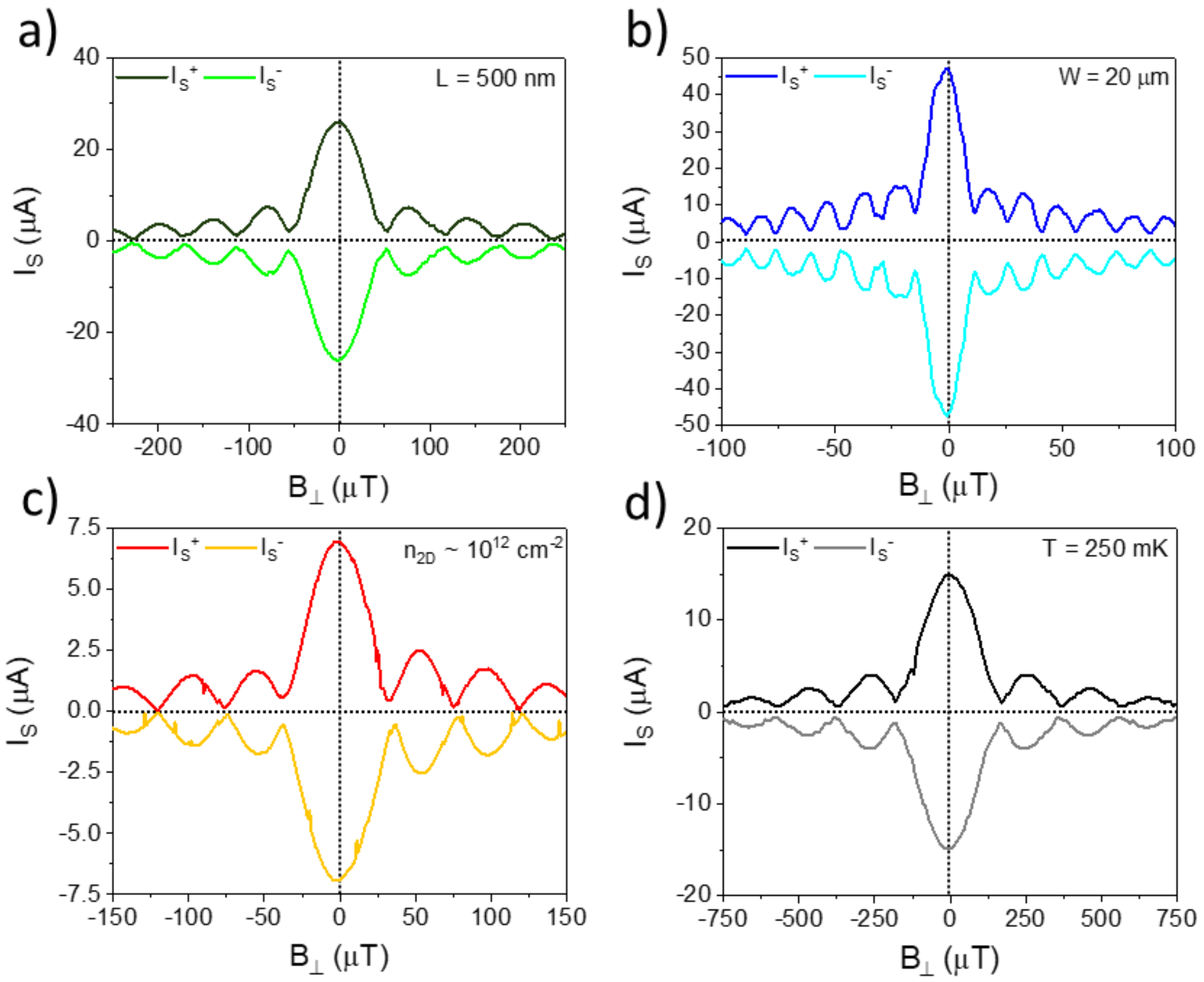

**Figure S10: Fraunhofer diffraction patterns of the fabricated Josephson Junctions.** a,b,c,d) Switching current vs. out-of-plane magnetic field for JJs with different lengths (a), widths (b), InAsOI doping level (c), and temperatures (d). This figure extends Figure 4 of the manuscript. JJs specific properties are reported in Figure 4.

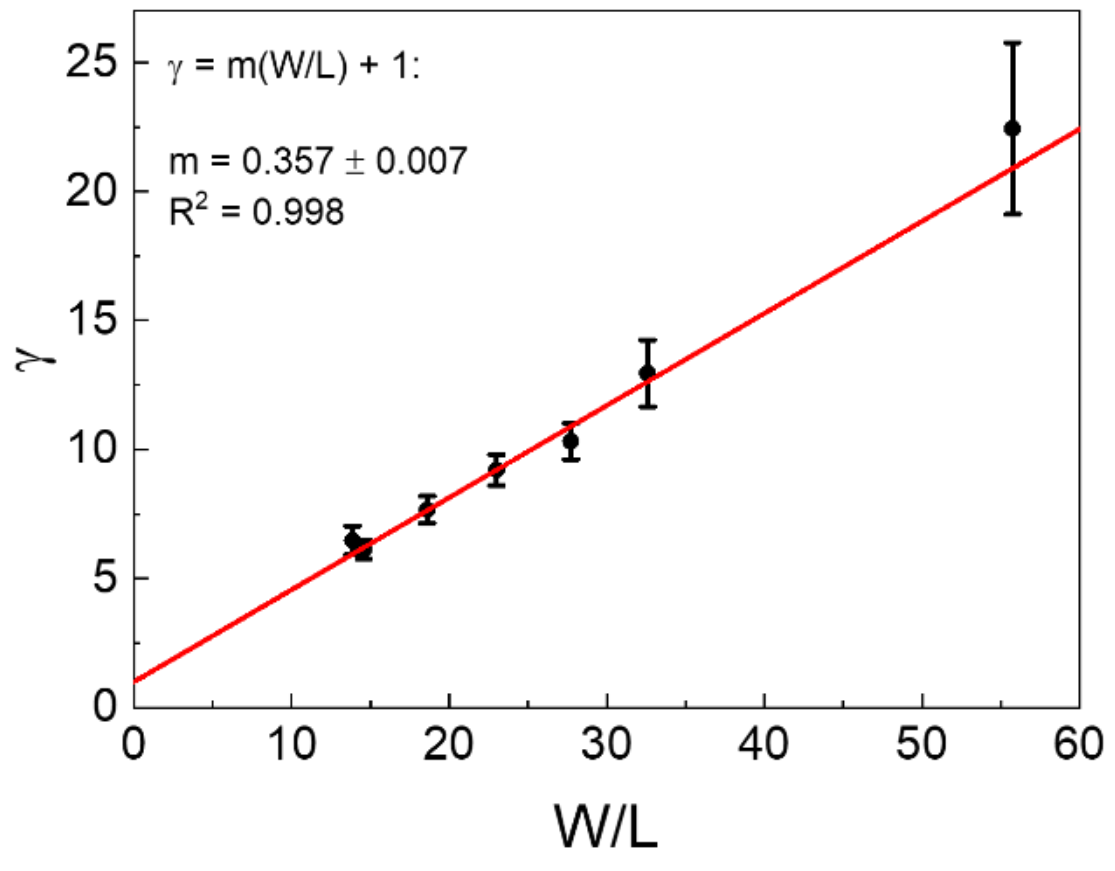

**Figure S11: InAsOI-based Josephson Junctions scaling factor (γ) as a function of aspect ratio (W/L). γ takes into account the magnetic field focusing and the enlargement of the magnetic effective JJ length compared to the inter-electrode separation.** The scaling factor is estimated as the ratio of the theoretically expected periodicity of the Fraunhofer pattern, given by $\Phi_0/A$ to the measured Fraunhofer pattern periodicity (p), expressed as $\gamma = \Phi_0/pA$. The periodicity is determined by sampling the minima of the Fraunhofer patterns and calculating the mean and standard deviation of the finite differences. We considered a set of Fraunhofer patterns with varying inter-electrode spacings, junction widths, and doping levels. The linear fit shows that the scaling factor depends linearly on the geometric aspect ratio of the JJs, increasing as the width is enlarged compared to the inter-electrode spacing. The intercept of the fit is compatible with $\gamma = 1$, indicating no flux focusing, as expected for narrow JJs. Data are reported as the average value measured over $n \geq 6$ minima for each W/L ratio, with error bars representing the standard deviation.

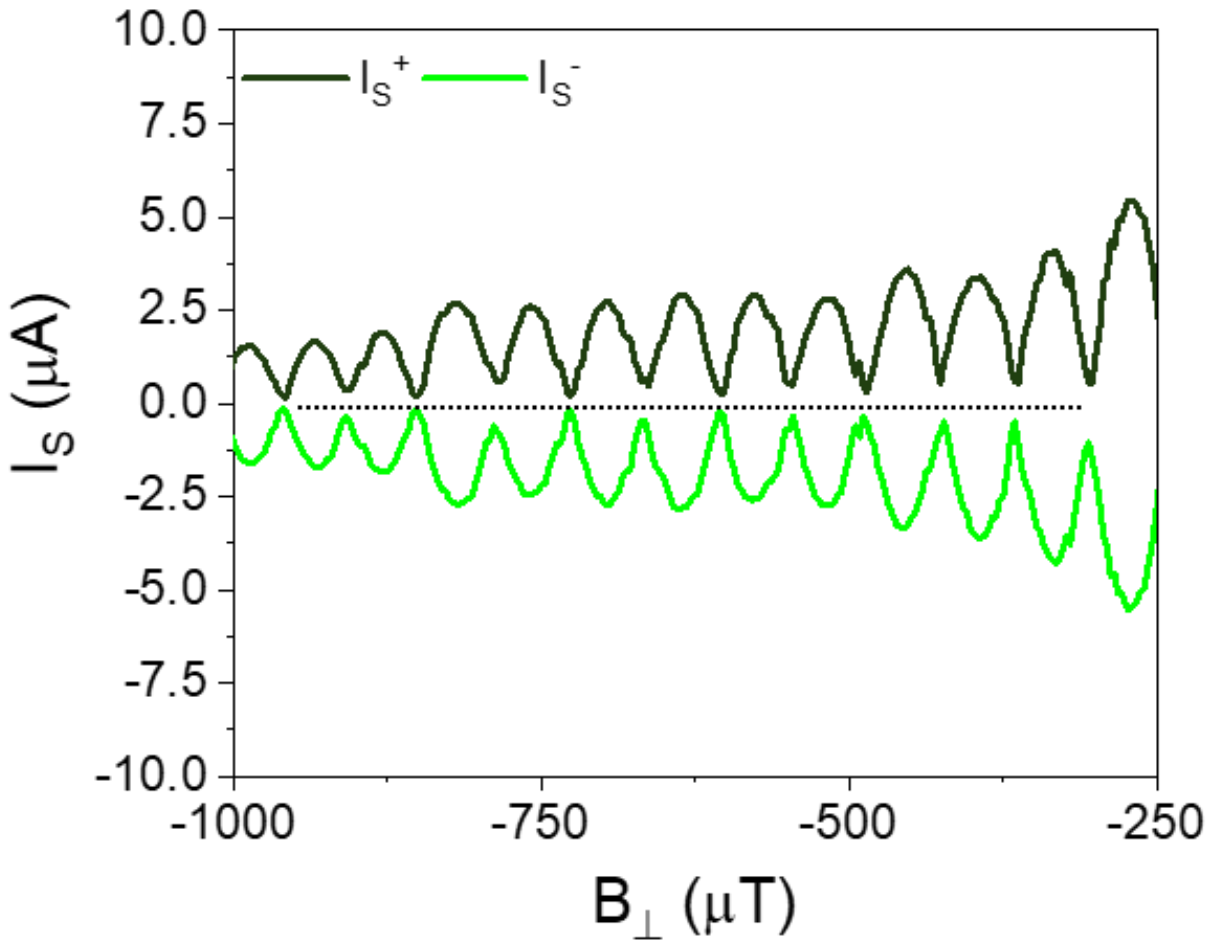

**Figure S12: Extended Fraunhofer diffraction pattern of a Josephson Junction with L = 350 nm, W = 10 µm, and $n_{2D} \sim 10^{13}$cm$^{-2}$.** The pattern minima approach values near 0 µA for B < -250 µT.

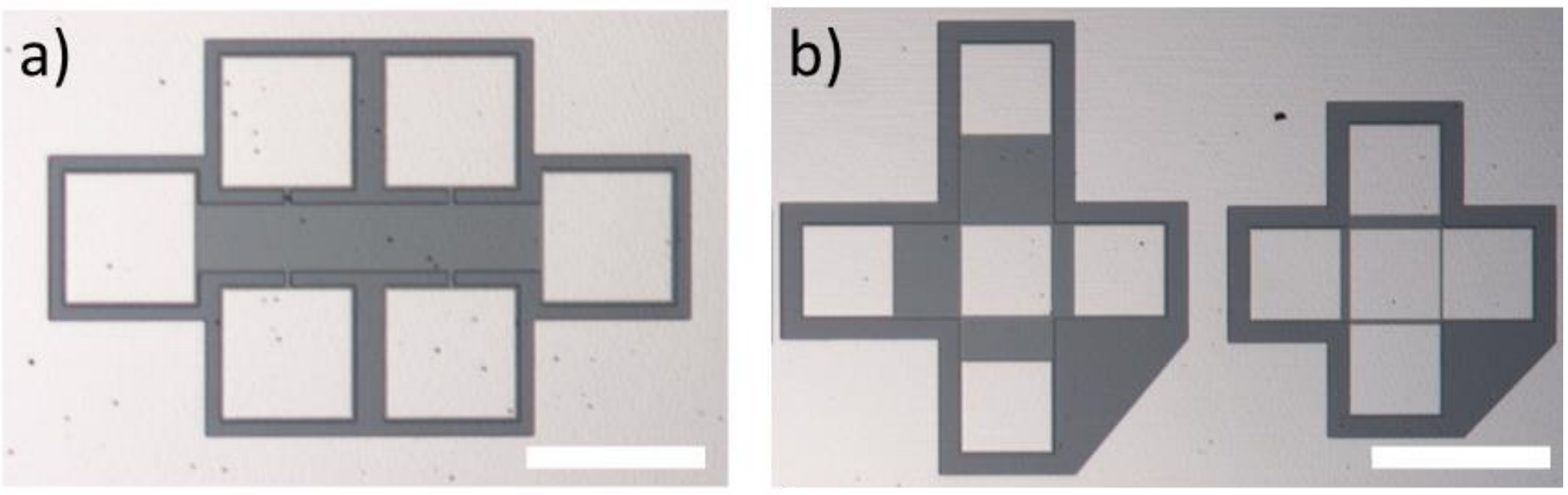

**Figure S13: Optical microscope photographs of 6-terminal Hall bars and TLM structures used to characterize the InAsOIs.** a) 6-terminal Hall bars geometry fabricated to estimate the room temperature and cryogenic resistivity, mobility, and sheet carrier density. Al pads (100 nm-thick) and the InAs bar (100 nm-thick) are isolated from the rest of the substrate by etching an InAs frame (100 nm-depth) all over the structure. b) TLM geometry fabricated to estimate the room temperature and cryogenic contact resistance. The electrical measurement was performed from the central pad to the outsider pads. Al pads (100 nm-thick) and the InAs lengths (100 nm-thick, 5 μm-to-200 μm length, 200 μm width) are isolated from the rest of the substrate by etching an InAs frame (100 nm-depth) all over the structure. The scalebar is 250 μm.

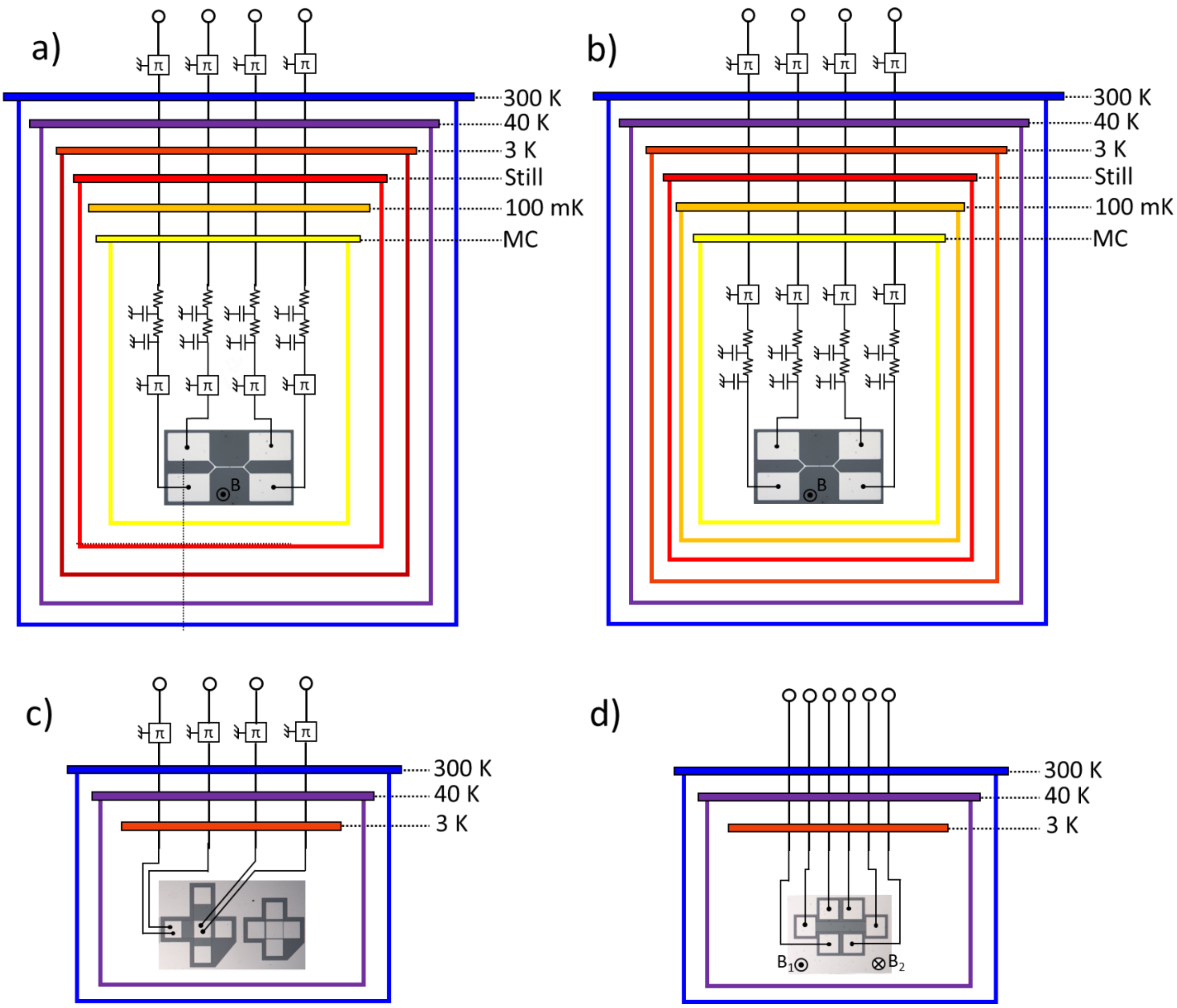

**Figure S14: Schematic view of the cryostat setups.** a,b) Schematic views of the Oxford Triton 200 (a) and Leiden CF-CS81-1400 (b) setups used to measure the JJs. c.d) Schematic views of the Ice Oxford DRY ICE 3K setups used with the TLM (c) and Hall bars (d) measurements. The resistance values are one kΩ, the capacitor values are 47 nF (a) and 4.7 nF (b), and the π-filter is OXLEY FLT/P/1500. The magnetic field is applied in agreement with the reported direction.

## 3. Materials and Manufacturing Methods

### 3.1 Materials and Chemicals

GaAs wafers (2’’ diameter, (100) orientation, $\rho=5.9\times10^7$ $\Omega\times$cm) were purchased from Wafer Technology LTD. Materials used for the Molecular Beam Epitaxy growth (Gallium 7N5, Aluminum 6N5, Indium 7N5, and Arsenic 7N5) were purchased from Azelis S.A.. Acetone (ACE, ULSI electric grade, MicroChemicals), 2-propanol (IPA, ULSI electric grade, MicroChemicals), S1805 G2 Positive Photoresist (S1805, Microposit, positive photoresist), AR-P 679.04 (AllResist, positive e-beam resist), MF319 Developer (MF319, Microposit), AR 600-56 Developer (AR 600-56, AllResist), AR600-71 (AllResist, remover for photo- and e-beam resist), Aluminum Etchant Type D (Transene), Phosphoric acid ($H_3PO_4$, Sigma Aldrich, semiconductor grade ≥85% in water), Hydrogen peroxide ($H_2O_2$, Carlo Erba Reagents, RSE-For electronic use-Stabilized, 30% in water), Nitrogen ($N_2$, 5.0, Nippon Gases) was provided by the Clean Room Facility of the National Enterprise for nanoScience and nanotechnology (NEST, Pisa, Italy). Diammonium sulfide ($(NH_4)_2S$, Carlo Erba Reagents, 20% in water) was provided by the Chemical Lab Facility of the NEST. Sulfur pieces (S, Alfa Aesar, 99.999% pure) was purchased from Carlo Erba Reagents S.r.l. Aluminum pellets (99.999% pure) were purchased from Kurt J. Lesker Company. Aqueous solutions were prepared using deionized water (DIW, 15.0 M$\Omega\times$cm) filtered by Elix® (Merck Millipore) provided by the Clean Room Facility of the NEST.

### 3.2 InAsOI Heterostructure Growth via Molecular Beam Epitaxy

InAsOI was grown on semi-insulating GaAs (100) substrates using solid-source Molecular Beam Epitaxy (MBE, Compact 21 DZ, Riber). Starting from the GaAs substrate, the sequence of the layer structure includes a 200 nm-thick GaAs layer, a 200 nm-thick GaAs/$Al_{0.16}Ga_{0.84}As$ superlattice, a 200 nm-thick GaAs layer, a 1.250 µm-thick step-graded $In_XAl_{1-X}As$ metamorphic buffer (with X increasing from 0.15 to 0.81), a 400 nm-thick $In_{0.84}Al_{0.16}As$ overshoot layer, and a 100 nm-thick InAs layer. The GaAs layer and the GaAs/$Al_{0.16}Ga_{0.84}As$ superlattice below the $In_XAl_{1-X}As$ buffer layer are grown to planarize the starting GaAs surface and to reduce surface roughness caused by the oxide desorption process. Both are grown with a group V/III beam flux ratio of 6.

The metamorphic buffer consists of two regions with different misfit gradients. The first $In_XAl_{1-X}As$ region is composed of twelve 50 nm-thick layers with X ramping from 0.15 to 0.58. The second $In_XAl_{1-X}As$ region is composed of twelve 50 nm-thick layers with X ramping from 0.58 to 0.81. The Al flux was kept constant during the buffer layer growth, while the In flux was increased at each step without growth interruptions. At the end of the buffer, the overshoot layer was grown to increase the strain relaxation of the $In_XAl_{1-x}As$ metamorphic layer [37]. The As flux was adjusted during the growth of the metamorphic buffer and the overshoot layer to keep a constant group V/III beam flux ratio of 8. The InAs layer was growth with a group V/III beam flux ratio of 8 and a growth rate of 0.96 μm/h. The GaAs layer and GaAs/AlGaAs superlattice are grown at 600 °C ± 5 °C. The metamorphic buffer and the overshoot layers were grown at optimized substrate temperatures of 320 °C ± 5 °C. The InAs epilayer was grown at 480 ± 5 °C. The doping of the InAs layer is achieved by using Si cell at different temperatures, namely 1230 and 1300 °C.

### 3.3 InAsOI Josephson Junction Fabrication

All wetting steps were performed in cleaned glass Beckers using stainless steel tweezers provided with carbon tips. Teflon-coated tweezers were used for all the steps requiring acid or base solutions. The Julabo TW2 was used to head the solution to a specific temperature.
InAsOI substrates were cut in square samples (7×7 mm×mm) and sonicated (Transonic, T310/H) in ACE and IPA for 5 min to remove GaAs dusts. The InAs air-exposed surface was etched from native InAs oxide ($InAsO_X$) and passivated with S-termination dipping the InAsOI samples in a $(NH_4)_2S_X$ solution (290 mM $(NH_4)_2S$ and 300 mM S in DIW) at 45°C for 90 s. The S-terminated InAsOI samples were then rinsed twice in DIW for 30 s and immediately loaded (~ 90 s exposure time in the air) into the load-lock vacuum chamber of an e-beam evaporator (acceleration voltage 7 kV). Samples were transferred into the deposition chamber, where a 100-nm-thick Al layer was deposited at a rate of 2 A/s at a residual chamber pressure of 1E-8 ÷ 5E-9 Torr. After Al deposition, a layer of S1805 positive photoresist was spin-coated at 5000 RPM for 60 s (Laurell Technologies, WS-650SZ-6NPP/LITE, spin coating acceleration of 5000 RPM/s) and soft-baked at 115 °C for 60 s (ATV Technologie, HT-304). The resist was then exposed via direct writing UV lithography (DMO, ML3 laser writer, λ=385 nm) with a dose of 60 $mJcm^{-2}$, resolution of 0.6 μm, high exposure quality, and laser-assisted real-time focus correction to define the MESA geometry. Unless otherwise stated, all the rinsing steps were performed at room temperature (RT, 21 °C). The UV-exposed samples were developed in MF319 for 45 s with soft agitation to remove exposed photoresist, then rinsed in DIW for 30 s to stop the development and dried with $N_2$. The exposed Al layer was removed by dipping the sample in Al Etchant Type D at 40 °C for 65 seconds with soft agitation, then rinsed in DIW for 30 seconds to stop the etching and dried with $N_2$. The exposed InAs epilayer was etched by dipping the samples in a $H_3PO_4$:$H_2O_2$ solution (348 mM $H_3PO_4$, 305 mM $H_2O_2$ in DIW) for 60 s with soft agitation, then rinsed in DIW for 30 s to stop the etching and dried with $N_2$. Eventually, the photoresist was removed by rinsing the InAsOI samples in ACE at 60 °C for 5 minutes and IPA for 60 s, then dried with $N_2$. At the end of this step, we achieved a Josephson Junction width (W) of 5, 10, or 20 μm. After MESA fabrication, a layer of AR-P 679.04 positive e-beam resist was spin-coated at 4000 RPM for 60 s (Laurell Technologies, WS-650SZ-6NPP/LITE, spin coating acceleration of 10000 RPM/s) and soft-baked at 160 °C for 60 s (ATV Technologie, HT-304). The resist was then exposed via marker-aligned e-beam lithography (ZEISS, Ultra Plus) with a dose of 350 $\mu Ccm^{-2}$, voltage acceleration of 30 kV, aperture of 7.5 μm, and line step size of 1 nm to define the JJ length (L). The electron-exposed samples were developed in AR 600-56 for 90 s with soft agitation to remove the exposed e-beam resist, then rinsed in IPA for 30 s to stop the development and dried with $N_2$. Subsequently, the exposed Al layer was removed by dipping the sample in Al Etchant Type D at 40 °C for 65 seconds with soft agitation, then rinsed in DIW for 30 seconds to stop the etching and dried with $N_2$. Eventually, the e-beam resist was removed by rinsing the InAsOI samples in ACE at 60 °C for 5 min, IPA for 60 s, and dried with $N_2$. At the end of this step, we achieved a Josephson Junction length of 350, 500, 700, or 900 nm.

### 3.4 InAsOI Hall Bars and Transmission Lines Fabrication

All the wetting steps were performed in cleaned glass Beckers using stainless steel tweezers provided with carbon tips. Teflon-coated tweezers were used for all the steps requiring acid or base solutions. The Julabo TW2 was used to heat the solution to a specific temperature.
InAsOI substrates were cut into square samples (7×7 mm×mm) and sonicated in ACE and IPA for 5 min to remove GaAs dusts. The air-exposed InAs surface was etched from native InAs oxide ($InAsO_X$) and passivated with S-termination by dipping the InAsOI samples in a $(NH_4)_2S_X$ solution (290 mM $(NH_4)_2S$ and 300 mM S in DIW) at 45°C for 90 s. The S-terminated InAsOI samples were then rinsed twice in DIW for 30 s and immediately loaded (~ 90 s exposure time in the air) into the load-lock vacuum chamber of an e-beam evaporator (acceleration voltage 7 kV). Samples were transferred into the deposition chamber, where a 100-nm-thick Al layer was deposited at a rate of 2 A/s at a residual chamber pressure of 1E-8 ÷ 5E-9 Torr. After Al deposition, a layer of S1805 positive photoresist was spin-coated at 5000 RPM for 60 s (Laurell Technologies, WS-650SZ-6NPP/LITE, spin coating acceleration of 5000 RPM/s) and soft-baked at 115 °C for 60 s (ATV Technologie, HT-304). Then, the photoresist was exposed via direct writing UV lithography (DMO ML3 laser writer, $\lambda$=385 nm) with a dose of 60 $mJcm^{-2}$, resolution of 0.6 μm, high exposure quality, and laser-assisted real-time focus correction to define the Al geometry for fabrication of 6-terminals Hall bars and Transfer Lengths suitable for the Transfer Length Method (TLM). Unless stated otherwise, all the rinsing steps were performed at room temperature (RT, 21 °C). The UV-exposed samples were developed in MF319 for 45 s with soft agitation to remove exposed photoresist, then rinsed in DIW for 30 s to stop the development and dried with $N_2$. The exposed Al layer was removed by dipping the sample in Al Etchant Type D at 40 °C for 65 seconds with soft agitation, then rinsed in DIW for 30 seconds to stop the etching and dried with $N_2$. Then:

- To evaluate the InAlAs properties, the exposed InAs epilayer was etched by dipping the samples in a H3PO4:H2O2 solution (348 mM H3PO4, 305 mM H2O2 in DIW) for 60 s with soft agitation. The samples were then rinsed in DIW for 30 seconds to stop the etching and dried with $N_2$.
- To evaluate the InAs properties, no InAs epilayer etching was performed.

Eventually, the photoresist was removed by rinsing the InAsOI samples in ACE at 60 °C for 5 min and IPA for 60 s, which was then dried with $N_2$.

A layer of S1805 positive photoresist was spin-coated at 5000 RPM for 60 s (Laurell Technologies, WS-650SZ-6NPP/LITE, spin coating acceleration of 5000 RPM/s) and soft-baked at 115 °C for 60 s (ATV Technologie, HT-304). Then, the photoresist was exposed via a second marker-aligned direct writing UV lithography (DMO ML3 laser writer, $\lambda$=385 nm) with a dose of 60 mJcm$^{-2}$, resolution of 0.6 μm, high exposure quality, and no real-time focus correction, to define the InAs/InAlAs geometry for fabrication of Hall bars and Transfer Lengths suitable for the TLM. The UV-exposed samples were developed in MF319 for 45 s with soft agitation to remove exposed photoresist, then rinsed in DIW for 30 s to stop the development and dried with $N_2$. Then:

- To evaluate the InAlAs properties, all the MBE-growth heterostructure was etched to expose the GaAs substrate by dipping the samples in a $H_3PO_4$:$H_2O_2$ solution (348 mM $H_3PO_4$, 305 mM $H_2O_2$ in DIW) for 14 min and 30 s with soft agitation, then rinsed in DIW for 30 s to stop the etching and dried with $N_2$.
- To evaluate the InAs properties, the exposed InAs epilayer was etched by dipping the samples in a $H_3PO_4$:$H_2O_2$ solution (348 mM $H_3PO_4$, 305 mM $H_2O_2$ in DIW) for 60 s with soft agitation, then rinsed in DIW for 30 s to stop the etching and dried with $N_2$.

Eventually, the photoresist was removed by rinsing the InAsOI samples in ACE at 60 °C for 5 min, IPA for 60 s, and drying with $N_2$. We fabricated 6-contact Hall bars with a width of 100 μm, source-to-drain length of 500 μm, and probe-to-probe length of 250 μm (Figure S13a). We fabricated Transfer Lengths for the TLM with a width of 200 μm and lengths of 5, 10, 20, 30, 50, 100, 150, and 200 μm (Figure S13b).

### 3.5 Sample Bonding via Wire Wedge Bonding

All the fabricated samples were provided with bonding pads ranging from 150 × 150 to 200 × 200 μm×μm and then used to connect the device with the chip carrier. Samples were glued using a small drop of AR-P 679.04, then left dry at RT for 1 hour on a 24-pin dual-in-line (DIL) chip carrier. Samples were bonded via wire wedge bonding (MP iBond5000 Wedge) using an Al/Si wire (1%, 25 μm wire diameter), leaving the user-bonder and the DIL chip carrier electrically connected to the ground.

# 4. Characterization Methods

## 4.1 Morphological Characterization

***via Scanning Electron Microscopy***

Top view morphological characterization of JJs was carried out via scanning electron microscopy (SEM, ZEISS Merlin) with 5 kV acceleration voltage, 178 pA filament current, back scattered electron relevator, at different magnifications (2.5k and 50k).

***via Optical Microscopy***

Optical microscopy (Leica, DM8000 M, provided with LEICA MC190 HD camera) was used to verify all the steps without photoresist. An optical microscope (Nikon, Eclipse ME600, provided with Nikon TV Lens C-0.6× and a UV filter) was used to evaluate all the steps involving the photoresist.

***via Atomic Force Microscopy***

Atomic force microscopy (AFM, Bruker, DIMENSION edge with ScanAsyst provided with an ASYLEC-01-R2 tip - silicon tip Ti/Ir coated, $f_0$=75 kHz, k=2.8 N/m - in tapping mode) was used to measure RMS roughness values of InAs epilayers, InAlAs metamorphic buffer layer, and Al thin film deposited. AFM analysis was also performed to evaluate the Josephson Junction shape. All the AFM photos were processed using Gwyddion.

***via Stylus Profilometry***

Stylus profilometry (Bruker, DektakXT, stylus radius 12.5 μm, stylus force 3 mg) was used to evaluate the thickness achieved after all the etching processes.

## 4.2 Electrical Characterization

### 4.2.1 Cryogenic and Room Temperature DC Electrical Characterization of InAsOI Transport Properties

Electrical characterization of InAs epilayers and InAlAs metamorphic buffer layers was carried out by measuring resistivity and mobility via 6-terminals Hall bars (HBs) geometry and contact resistance via the TLM, both at 300 K and 3 K. For the contact resistance measurement, samples were mounted in contact with the 3 K plate of the Ice Oxford DRY ICE 3K cryostat. The electrical and thermal configurations of the cryostat are shown in Figure S14c. 4-wire measurements are employed to evaluate the 4-wire resistance of each contact length used in the TLM. A current sweep from 0 to 1 μA (KEITHLEY 2400 SourcMeter, 50 nA step size) was applied while the voltage drop across the contacts was read (KEITHLEY 2400 SourcMeter, NPLC = 2; or HP 34401 Multimeter, NPLC = 2). The contact resistance was evaluated by dividing by two the y-axis intercept of the linear best-fitting resistance vs. length curve, while the transmission length was assessed by dividing by two the x-axis intercept of the linear best-fitting resistance vs. length curve (Figure S5).

Due to the insulating behavior of the InAlAs metamorphic buffer layer at temperatures lower than 70 K, I-V curves of each contact length employed in the TLM were evaluated by 2-wire measurements by applying a voltage sweep from 0 to 30 V (YOKOGAWA GS200 DC Voltage/Current Source, 250 mV step, 1 s waiting time point-to-point) and collecting the amplified current (FEMTO DDPCA-300, A=1E11, full bandwidth/rising time fast) flowing from the device (HP 34401 Multimeter, NPLC = 2).

The InAlAs resistance vs. temperature behavior was evaluated by 2-wire measurements, dividing the applied voltage drop of 1 V (KEITHLEY 2400 SourcMeter) by the collected amplified current (FEMTO DDPCA-300, A=1E10÷1E7÷1E4, full bandwidth/rising time fast; HP 34401 Multimeter, NPLC = 2) flowing from the five μm-contact-length device employed in the TLM. The cryostat was warmed up from 3 K to 300 K during the electrical measurement.

Hall measurements were performed on 6-terminal HBs to determine the InAs sheet electron density, mobility, and resistivity. Samples were mounted in contact with the 3 K plate of the Ice Oxford DRY ICE 3K cryostat, which was provided with a permanent magnet (B=250 mT). The cryostat's electrical and thermal configurations are shown in Figure S14d. Two lock-in amplifiers (Stanford Research System SR830) were used to inject AC current and measure AC voltages.

Hall measurements were performed on HBs with a standard lock-in-amplifier-based technique. The first lock-in amplifier oscillator voltage ($V_{OSC,RMS}$ = 1 V, f = 13.321 Hz) was applied across a series resistor (R=10 MΩ) at least 100 times larger than the total resistance of the remaining measurement setup to use the AC current ($I_{OSC,\ RMS}$). The AC current is injected into the sample's source contact. In contrast, the flowing current is measured with the drain contact, and the second lock-in amplifier measures the voltage drop's magnitude and phase across two other contacts. The contacts are manually switched to inject the current and measure the voltage. The Hall measurements were performed in three steps. In the first step, the resistivity ($\rho$) is calculated at zero magnetic fields using the formula $\rho = R_{xx} \times \frac{W \times t}{l}$, where $R_{xx}$ is the resistance measured between longitudinal (same-side) contacts of the HB for a current passing between source and drain, *W* is the HB width, *t* is the InAs thickness, and *l* is the distance between probe contacts. In the second and third steps, the Hall voltage ($V_H$) is measured by applying positive (Figure S14d, $B_1$=B) and negative (Figure S14d, B2=B) out-of-plane magnetic fields, respectively. $V_H$ is measured between opposite probe contacts of the HB for a current passing through the source and drain. The sheet electron density ($n_{2D}$) is calculated using the formula $n_{2D} = \frac{n_{2D}^1 + n_{2D}^2}{2}$, with $n_{2D}^i = \frac{I_{OSC,RMS} \times B_i}{q \times V_H}$, where *q* is the fundamental charge, and $B_i$ is the magnitude of the applied out-of-plane magnetic field. The electron mobility ($\mu_n$) is calculated using the formula $\mu_n = \frac{\mu_n^1 + \mu_n^2}{2}$, with $\mu_n^i = \frac{1}{n_{2D}^i \times q \times R_{xx}}$.

#### 4.2.2 Cryogenic and Room Temperature Electrical Characterization of InAsOI Josephson Junctions

Electrical characterization of Al-InAs JJs was carried out by measuring 4-wires I-V curves at different temperatures (ranging from 50 mK to 300 K) and different magnitudes of the out-of-plane magnetic field ($B_{\perp}$, ranging from -1 mT to 1 mT). The sample was mounted in contact with the mixing chamber (MC) plate of the Oxford Triton 200 or Leiden CF-CS81-1400 cryostat. Electrical and thermal configurations of the cryostats are shown in Figure S14a,b. Forward and backward source-drain current sweeps were applied using a voltage drop (YOKOGAWA GS200 DC Voltage/Current Source, step depending on the maximum current) across a series resistor (R=100 kΩ to 1 MΩ) at least 100 times larger than the total resistance of the remaining measurement setup. The voltage drop across the probe contacts was amplified (DL Instruments 1201, Gain = 10k, High pass filter = DC, Low pass filter = 100 Hz or 30 Hz) and read (34401 Multimeter, NPLC = 2 or 0.6). The switching current ($I_S$) was estimated as the last applied current in the upward curve before reading a voltage drop different from the noise floor. The re-trapping current ($I_R$) was calculated as the previously applied current in the downward curve before reading a voltage drop in the noise floor.

For the acquisition of the $I_S$ vs. $B_{\perp}$ interference pattern, the procedure was automated using an acquisition program capable of performing a second-order derivative of the probe voltage with respect to the source-drain current in real-time. By setting a voltage threshold, the program recognizes the transition from the dissipation-less to the dissipative regime in the calculated differential resistance and determines $I_S$. The normal state resistance was estimated as the slope of the I-V curve applying currents larger than the switching current.

The resistance vs. temperature (R vs. T) behavior was acquired by measuring the differential resistance of the JJ during the cryostat cool down starting from 1.5 K. A digital lock-in amplifier (NF Corporation, LI 5640, oscillator voltage RMS = 0.05 V, oscillator voltage frequency = 17 Hz, integration time = 100 ms, sensitivity = 100 mV) was used to provide an AC source-drain current and measure the differential resistance of the JJ. The lock-in amplifier oscillator voltage was applied across a series resistor (R=100 kΩ) at least 100 times larger than the total resistance of the remaining measurement setup to apply the AC source-drain current. The AC voltage drop across the probe contacts was amplified (DL Instruments 1201, Gain = 10k, High pass filter = DC, Low pass filter = 100 Hz or 30 Hz) and read by the lock-in amplifier to extrapolate the JJ differential resistance. Then, the data output voltage of the lock-in amplifier has been read (34401 Multimeter, NPLC = 0.1) and recorded.

## 5. Theoretical Analysis

### 5.1 InAs Coherence Length Calculation

The InAs coherence length ($\xi_N$) is calculated using the following equation [38]:

$$(1)\ \xi_N(T) = \sqrt{{hD}/{4\pi k_B T}}$$

where $D$ is the diffusion constant of the InAs epilayer, $T$ is the electron temperature, $h$, and $k_B$ are the Planck and Boltzmann constant, respectively.

$D$ is calculated from the relationship:

$$(2)\ D = {1}/{e^2 \rho n_F}$$

where $\rho$ is the resistivity of the InAs layer, and $n_F$ is its density of states at the Fermi level. $n_F$ is extrapolated from the formula:

$$(3)\ n_F = {(8\pi^2 m^*/h^2)^{3/2}\sqrt{E_F}}/{2\pi^2}$$

where $m^* = 0.023 m_e$ is the effective mass of InAs with $m_e$ the electron mass, and $E_F$ is the Fermi energy. The Fermi energy depends on the bulk electron density ($n_{3D} = n_{2D}/t$) via the relationship:

$$(4)\ E_F = \left({h^2}/{8\pi^2 m^*}\right)(3\pi^2 n_{3D})^{2/3}$$

we assume that the donors are consistently fully ionized in the conduction band together with the Fermi level positioning within the conduction band. We also assume a uniform charge density distribution over the InAs thickness.

Another definition of the coherence length ($\xi_0$) in an SNS junction, relevant for modeling the vortex state within the SNS JJ and consequently its response to a magnetic field, is:

$$(5)\ \xi_0 = \sqrt{\hbar D/\Delta}$$

where $\Delta = 1.76\ k_B T_C$ is the superconducting gap of the aluminum banks, $k_B$ is the Boltzmann constant, and $T_C$ is the critical temperature [39][40].

### 5.2 InAs Electron Mean Free Path Calculation

The InAs electron mean free path ($l_e$) was estimated from the following relationship:

$$(6)\ l_e = {3D}/{v_F}$$

where $v_F = \sqrt{2E_F/m^*}$ is the Fermi velocity of the carriers.

## 5.3 Considerations on the Switching Current vs. Temperature Characteristics: Thouless Energy and Effective Length Estimation

The Thouless energy ($E_{Th}$) of an SNS junction represents the characteristic energy scale of electron diffusion in a finite conductor, and it is an essential parameter in describing the physics of a proximitized weak link [38]. $E_{Th}$ can be evaluated by analyzing the switching current vs. temperature behavior. In the long junction regime ($\Delta \gg E_{Th}$, where $\Delta$ is the BCS gap of the superconductor) and high device temperature ($k_B T \gg E_{Th}$), the mutual influence of the two superconducting leads can be neglected, allowing the Usadel equations to be linearized in the normal metal, except in the vicinity of the NS interfaces. This yields:

$$(7)\ I_S(T) = \frac{64\pi k_B T}{eR_N}\sum_{n=0}^{\infty}\frac{\sqrt{\frac{2\omega_n(T)}{E_{Th}}}\,\Delta^2(T)\,e^{\sqrt{\frac{2\omega_n(T)}{E_{Th}}}}}{\left\{\omega_n(T)+\Omega_n(T)+\sqrt{2\left[\Omega_n{}^2(T)+\omega_n(T)\Omega_n(T)\right]}\right\}^2}$$

where $\omega_n(T) = (2n+1)\pi k_B T$ are the Matsubara frequencies for fermions and $\Omega_n(T) = \sqrt{\Delta^2(T) + \omega_n{}^2(T)}$ [41]. By fitting this equation with $E_{Th}$ as a parameter, it is possible to extract the Thouless energy and, subsequently, the effective length ($L_{eff}$) of the SNS junction, which in the diffusive limit is given by:

$$(8)\ L_{eff} = \sqrt{\frac{hD}{2\pi E_{Th}}}$$

where D is given by (2).

While this model is commonly employed to estimate the physical properties of a long diffusive SNS junction, it does not apply to the Al/InAs junctions examined in this paper. The reason for the model's inadequacy in describing the physics of the proposed devices lies in the NS interface since the InAs epilayer spreads beneath the superconductive leads. Consequently, there are superconductive channels with different lengths through which the supercurrent can flow, resulting in a junction not strictly defined by the inter-electrode separation. The effect of a widely spread NS interface on an SNS junction $I_S$ vs. $T$ characteristic was already observed in other devices, specifically in ballistic SNS junctions, where the model proposed by the authors fails to represent the experimental data [42].

## 6. Statistical Analysis

*Pre-processing of data*

Critical current density was evaluated by dividing the measured critical current value by the JJ width assessed by SEM.

*Data presentation (e.g., mean ± SD)*

All the data referring to more than three devices are presented as mean value ± standard deviation.

*Sample size (n) for each statistical analysis*

All the data referring to more than three devices are presented, indicating the sample size (n).

*Software used for analysis*

Best fitting and statistics of experimental data were performed using ORIGIN 2023.